\documentclass[a4paper,UKenglish,cleveref, autoref, thm-restate]{lipics-v2021}

\title{Evaluating Optimal Safe Flow Decomposition for RNA Assembly}

\titlerunning{Evaluating Optimal Safe Flows Decomposition for RNA Assembly} 

\author{Bashar Ahmed}{Department of Computer Science and Engineering, Indian Institute of Technology Roorkee, India}{bashar_a@cs.iitr.ac.in}{}{}
                
\author{Siddharth Singh Rana}{Department of Computer Science and Engineering, Indian Institute of Technology Roorkee, India}{siddharth_sr@cs.iitr.ac.in}{}{}

\author{Ujjwal}{Department of Computer Science and Engineering, Indian Institute of Technology Roorkee, India}{ujjwal@cs.iitr.ac.in}{}{}
                
\author{Shahbaz Khan}{Department of Computer Science and Engineering, Indian Institute of Technology Roorkee, India}{shahbaz.khan@cs.iitr.ac.in}{https://orcid.org/0000-0001-9352-0088}{}

\authorrunning{B. Ahmed et al.} 

\Copyright{Bashar Ahmed, Siddharth Singh Rana, Ujjwal and Shahbaz Khan} 



\category{} 

\supplement{}

\funding{Startup Research Grant SRG/2022/000801 by DST SERB, Government of India.}

\acknowledgements{We would like to anonymous reviewers to highlight crucial issues with our previous submission regarding the scalability of our results.}

\nolinenumbers 

\hideLIPIcs  

\EventEditors{Shiri Chechik, Gonzalo Navarro, Eva Rotenberg, and Grzegorz Herman}
\EventNoEds{4}
\EventLongTitle{30th Annual European Symposium on Algorithms (ESA 2022)}
\EventShortTitle{WABI 2024}
\EventAcronym{WABI}
\EventYear{2024}
\EventDate{September 5--9, 2024}
\EventLocation{Berlin/Potsdam, Germany}
\EventLogo{}
\SeriesVolume{244}
\ArticleNo{88}

\usepackage{amsmath}
\usepackage{amssymb}
\PassOptionsToPackage{hidelinks}{hyperref}
\usepackage[dvipsnames]{xcolor}
\usepackage[linesnumbered,ruled,vlined]{algorithm2e} 
\usepackage{multicol}
\usepackage{multirow}
\usepackage{comment}

\usepackage[justification=centering]{caption}
\usepackage{subcaption}

\usepackage{graphicx}
\usepackage{array}
\usepackage{float}
\usepackage[normalem]{ulem}

\newcommand{\sbzkc}[1]{{\color{cyan} (#1)}}

\newcolumntype{P}[1]{>{\centering\arraybackslash}p{#1}}

\begin{document}

\maketitle

\begin{abstract}
Flow decomposition is a fundamental graph problem with numerous applications. In Bioinformatics, the applications of flow decomposition in directed acyclic graphs are prominently highlighted in the RNA Assembly problem. However, flow decomposition admits multiple solutions where exactly one solution correctly represents the underlying RNA transcripts. Further, no parameter exists that can be optimized to compute the required solution. The problem was thus addressed by Safe and Complete framework~[RECOMB16], which reports all the parts of the solution that are present in every possible solution, thus can be \textit{safely} reported with $100\%$ precision. 

Khan et al.~[RECOMB22] first studied the flow decomposition in the safe and complete framework. Their algorithm showed superior performance ($\approx20\%$) over the popular heuristic (greedy-width) on sufficiently complex graphs for a unified metric of precision and coverage (F-score). Moreover, it was also shown to have superior time and space efficiency. They presented multiple representations of the solution using simple but suboptimal algorithms. Later, optimal algorithms for the existing representations were presented in a theoretical study by Khan and Tomescu~[ESA22], who also presented an optimal representation requiring \textit{constant} space per safe path. 

In this paper, we evaluate the practical significance of the optimal algorithms by Khan and Tomescu~[ESA22]. Since all the evaluated algorithms are still safe and complete, the evaluation is restricted to time and space efficiency as the precision and coverage (and hence F-score) metrics remain the same. 
Our work highlights the significance of the theoretically optimal algorithms improving time (up to $60-70\%$) and memory (up to $76-85\%$), and the optimal representations improving output size (up to $135-170\%$) significantly. However, the impact of optimal representations on time and memory was limited due to a large number of extremely short safe paths. To address this limitation, we propose heuristics to improve these representations further, resulting in further improvement in time (up to $10\%$) and output size ($10-25\%$). However, in absolute terms, these improvements were limited to a few seconds on real datasets involved due to the smaller size of the graphs. We thus generated large random graphs having similar results as real graphs, to demonstrate the scalability of the above results. The older algorithms [RECOMB22] were not practical on moderately large graphs ($\geq 1M$ nodes), while optimal algorithms [ESA22] were linearly scalable for much larger graphs ($\geq 100M$ nodes). 

\end{abstract}

\newpage
\section{Introduction}

\setcounter{page}{1}
Network flow is among the most studied graph problems having important practical applications. 
For a directed acyclic graph having $n$ vertices and $m$ edges with a unique source $s$ and a unique sink $t$, its flow decomposition is a weighted set of $s$-$t$ paths such that the sum of weights of paths containing an edge is equal to the flow on the edge in the input flow graph~\cite{FordNetwork}. 
Prominent applications of flow decomposition range from transportation~\cite{hong2013achieving,cohen2014effect,mumey2015parity}, network routing~\cite{hong2013achieving,hartman2012split,mumey2015parity} to recent applications in bioinformatics~\cite{pertea2015stringtie,tomescu2013novel,gatter2019ryuto,bernard2013flipflop,TomescuGPRKM15,williams2019rna,DBLP:conf/recomb/BaaijensSS20,BaaijensRKSS19}. 


However, a given flow network admits multiple flow decompositions, requiring further parameters to formalize the optimal solution. For example, minimum flow decomposition reports the flow decomposition with the minimum number of decomposed paths, though it is NP-hard to compute even for DAGs~\cite{vatinlen2008simple}. This led to focus on approximate algorithms~\cite{hartman2012split,SUPPAKITPAISARN2016367,pienkosz2015integral,mumey2015parity,baier2005k} and practical heuristics~\cite{vatinlen2008simple,Shao}. For several applications in network routing and transportation, any flow decomposition is a valid solution. But in some applications, particularly in bioinformatics, exactly one of the possible decompositions is correct, while there is no parameter to be optimized that reports the correct solution.

Consider the example of RNA assembly problem, which computes the RNA transcripts from its short substrings and abundances (weights) found using high throughput sequencing~\cite{citeulike:3614773}. This data can be used to build a flow graph such that one of its flow decompositions gives the corresponding RNA transcripts with their abundances. However, no parameter can be optimized to report the flow decomposition corresponding to the RNA transcript. Hence, reporting any flow decomposition results in an incorrect solution. Such problems are often addressed using the \textit{safety} framework, which reports the part of the solution that is correct. 

\subsection{Problem Definition and Related Work}
The problem of multiple solutions to the flow decomposition was first addressed by Ma et al.~\cite{DBLP:conf/wabi/MaZK20} using a probabilistic framework. Thereafter, they presented a quadratic algorithm~\cite{findingranges} to verify if a path containing a given set of edges is present in every possible flow decomposition. Such a path will be present in the original RNA transcripts as it is a valid flow decomposition.

The safety framework was introduced by Tomescu and Medvedev~\cite{tomescu2017safe} for the genome assembly problem. For a problem admitting multiple solutions, a partial solution is said to be \textit{safe} if it appears in all solutions to a problem. They extend the notion to \textit{safe and complete}, which reports \textit{everything} that can be \textit{safely} reported. 
Later, several related works developed the same for problems computing edge or node covering problems of the given graph, under various settings as circular walks~\cite{CairoMART19,cairo2020macrotigs,acosta2018safe}, Eulerian walks~\cite{nagarajan2009parametric,acosta2024simplicity}, linear walks~\cite{cairo2020safety,Cairo0RSTZ23}.

A path is safe for the flow decomposition problem if it is a subpath of \textit{at least one} $s$-$t$ path in \textit{every} set of $s$-$t$ paths which is a valid flow decomposition of the input graph. 
Khan et al.~\cite{KhanMMLA22} first addressed the flow decomposition problem in the safe and complete framework. They presented a characterization of safe flow paths and presented an optimal verification algorithm to verify whether a given path is safe in the input flow graph. They further presented a simple enumeration algorithm for enumerating all the safe paths in $O(mn+|out|)$ time, where $out$ is the set of maximal safe paths. Their algorithm had a provision to present the output in two formats, namely \textit{raw} (reporting each safe path explicitly) and \textit{concise} (combining some overlapping safe paths to save space). 
They also proved the worst-case size of raw representation to be $\Omega(mn^2)$ and that of concise representation to be $\Omega(mn)$. They evaluated their algorithm with trivially safe algorithms (unitigs~\cite{KM95} and extended unitigs~\cite{PTW01}) and a popular heuristic (greedy-width~\cite{vatinlen2008simple}), reporting superior performance in both F-score (combined metric for precision and coverage) as well as time and space. 

Khan and Tomescu~\cite{optFlowTheory} further developed the theory of safe flow decomposition, presenting new properties and optimizing algorithms and representations. For both raw and concise representations, they improved the solution to $O(m+|out|)$, whose worst-case size was now proven to be tight bounds of $\Theta(mn^2)$ and $\Theta(mn)$, respectively. Further, their algorithm for concise representation computed the minimum sized concise representation (as multiple concise representations are possible). Finally, they presented an optimal representation of safe paths requiring constant space for every safe path and presented an $O(m+|out|\log n)$ time algorithm for computing the same. However, this work was purely theoretical in nature, unlike the previous \cite{KhanMMLA22}, and hence the practical relevance of the new algorithms was uncertain.



\subsection{Our results}
The main aim of our work was the experimental evaluation of the optimal algorithms and representations developed by Khan and Tomescu~\cite{optFlowTheory}. Our algorithms were evaluated on the updated datasets for the RNA assembly problem used by \cite{KhanMMLA22}. Our results are as follows.

\begin{enumerate}
    \item \textbf{Augmenting implementation details.} Despite the intuition behind the algorithms being clearly described, there were several challenges in the implementation. Several errors were corrected in the pseudocode for the algorithm for optimal concise representation. The algorithm for optimal representation lacked several details required to achieve the desired bound. In particular, in order to avoid generating duplicates among safe paths, additional criteria were developed to ensure uniqueness. 
    
    \item \textbf{Empirical evaluation of safe algorithms and representations.} Given the clear dominance of the safe and complete algorithm over other trivially safe algorithms and heuristics, we evaluated the new algorithms~\cite{optFlowTheory} only against the previous safe and complete algorithm~\cite{KhanMMLA22}. The analysis indicates that the new algorithms improve the time (up to $60-70\%$) and memory (up to $76-85\%$) of the previous algorithms since they are not dependent on the underlying candidate flow decomposition. Moreover, the optimal concise representation improves the concise representation significantly (up to $170\%$). Furthermore, the optimal representation significantly improves other representations' output size (up to $135\%$). However, the performance of algorithms strongly depended on the output size, where in most cases, the optimal algorithm for raw representation was better.
    
    \item \textbf{Heuristics to improve both representations and algorithms.} Deeper investigations revealed that most reported safe paths were extremely short, resulting in wasteful concise and optimal representations. An exhaustive analysis of various representations of safe paths resulted in heuristics for concise and optimal representations. These heuristics significantly impacted the output size ($10-25\%$) and time taken (up to $10\%$), resulting in state-of-the-art practical algorithms. 
    
    \item \textbf{Evaluation on randomly generated graphs.} We further evaluated the performance of the algorithms on uniformly random graphs and random power law graphs. Surprisingly, the optimal algorithm for raw representation was faster despite the impact of concise and optimal representation heuristics, which made their output size smaller than the raw representation. The reason was the insignificant output size in comparison to the input. We thus proposed improved random graphs, which resulted in longer, safe paths having comparable output sizes. This resulted in behaviors similar to that of real graphs. The evaluation of larger such graphs additionally demonstrated the scalability of our results, showing that only optimal algorithms were scalable for larger graphs ($\geq 1M$ nodes).
\end{enumerate}

\textbf{Outline of the paper.} We first describe in \Cref{sec:prelims} the basic notations and definitions required in our paper, and \Cref{sec:sota} describes the state-of-the-art algorithms for each representation along with their limitations. \Cref{sec:experiment-evaluation} describes the experimental setup and \Cref{sec:evaluation-Real} describes the evaluation on real datasets. \Cref{sec:heuristics} describes our proposed heuristics to improve the algorithms and representations and their impact on the performance. Further, in \Cref{sec:random}, we evaluate and analyze the performance on randomly generated graph instances. We finally conclude our paper by summarizing our contributions in \Cref{sec:conclusion}.

\section{Preliminary}
\label{sec:prelims}
Consider a directed acyclic flow graph $G=(V,E)$ having a set of vertices $V$ with $|V|=n$ and a set of edges $E$ with $|E|=m$, where assuming a connected graph, we have $m\geq n - 1$. Every edge $e(x,y)$ has a positive flow $f(e)$ or $f(x,y)$, and the total flow on the incoming (or outgoing) edges of a vertex $v$ is represented by $f_{in}(v)$ (or $f_{out}(v)$). Further, a node $v$ is called a source if $f_{in}(v)=0$ and a sink if $f_{out}(v)=0$. The set of sources and sinks are respectively represented by $Source(G)$ and $Sink(G)$. Each vertex $v\in V\setminus \{Source(G)\cup Sink(G)\}$ follows the \textit{conservation of flow} where $f_{in}(v)=f_{out}(v)$. For any path $p$ in the graph, $|p|$ denotes the number of edges in the path. 

For every vertex $u$, $f_{max}(\cdot,u)$ (or $f_{max}(u,\cdot)$) denote the maximum incoming (or outgoing) flow from the vertex $u$. Similarly, $e_{max}(\cdot,u)$ (or $e_{max}(u,\cdot)$) denotes the edge corresponding to the maximum incoming (or outgoing) flow and is called the maximum incoming (or outgoing) edge. Further, 
$v_{max}(\cdot,u)$ (or $v_{max}(u,\cdot)$) denotes the other endpoint of the maximum incoming (or outgoing) edge and is called the maximum incoming (or outgoing) vertex. If the maximum incoming (or outgoing) edge is unique, we call it the \textit{unique} maximum incoming (or outgoing) edge and represent it as $e^*_{max}(\cdot,u)$ (or $e^*_{max}(u,\cdot)$), otherwise, the edge whose endpoint (other than $u$) comes earlier in the topological order is considered and called the \textit{preferred} maximum incoming (or outgoing) edge. Similarly, the other endpoint of the unique maximum incoming (or outgoing) edge is called the \textit{unique} maximum incoming (or outgoing) vertex and is represented as $v^*_{max}(\cdot,u)$ (or $v^*_{max}(u,\cdot)$), otherwise the endpoint corresponding to the preferred maximum incoming (or outgoing) edge is called as the \textit{preferred} maximum incoming (or outgoing) vertex. 
For a path $p$, the \textit{left extension} is done by prepending the unique maximum incoming edges and the \textit{right extension} is done by appending the unique maximum outgoing edges to the path. 



\begin{figure}[t]
    \centering
\includegraphics[trim=5 7 10 7, clip,scale=.8]{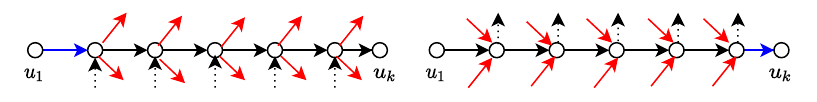}
    \caption{The excess flow of a path $P$ is the incoming  (or outgoing) flow [blue] that necessarily passes through the whole $P$ despite the flow [red] leaving (or entering) $P$ at its internal vertices (reproduced from \cite{KhanMMLA22}).}
    \label{fig:excessP}
\end{figure}

We further use the following notations for a safe path $p$ similar to ~\cite{KhanMMLA22}.
A path $P$ is called \textit{$w$-safe} if, in every possible flow decomposition, $P$ is a subpath of some paths in $\cal P$ (set of weighted paths) whose total weight is at least $w$. A $w$-safe path with $w>0$ is called a safe flow path or simply a safe path. A path $P$ is $w$-safe for any $w\leq f_P$ where $f_P$ is its excess flow of the path $P$ defined as follows (see \Cref{fig:excessP}).

\begin{definition}[Excess flow~\cite{KhanMMLA22}]
For a path $P=\{u_1,u_2,...,u_k\}$ we have 
\begin{equation*}
f_P= f(u_1,u_2) - \sum_{\substack{u_i\in \{u_2,...,u_{k-1}\} \\ 
v\neq u_{i+1}}} f(u_i,v)
= f(u_{k-1},u_k) - \sum_{\substack{u_i\in \{u_2,...,u_{k-1}\} \\ 
v\neq u_{i-1}}} f(v,u_i) \end{equation*}
\label{def:dispConv}
\end{definition}

A safe path is \textit{left maximal} (or \textit{right maximal}) if extending it to the left (or right) with any edge makes it unsafe. A safe path is \textit{maximal} if it is both left and right maximal. Further, we use the following property.

\begin{lemma}[Extension Lemma~\cite{KhanMMLA22}]\label{lem:excess-flow}
    For any path in a flow graph (DAG), adding an edge $(u, v)$ to its start or its end reduces its
    excess flow by $f_{in}(v) - f(u,v)$, or $f_{out}(u)-f(u,v)$, respectively.
\end{lemma}

\noindent\textbf{Note.}  A certain class of DAGs have a unique flow decomposition, and hence all the source-to-sink paths are naturally safe (and hence trivially
complete). 
This class of graphs is referred to as \textit{funnel instances}. A funnel graph thus becomes a trivial input for the problem as the entire decomposition is reported as safe.

\subsection{Representations of Safe Flow Decomposition}\label{subsection:repr}
\begin{figure}[h]
    \centering
    \includegraphics[scale=.9]{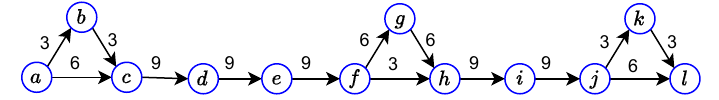}
    \caption{Example flow graph to describe safe path representations.}
    \label{fig:baseGraph}
\end{figure}

Consider \Cref{fig:baseGraph}, the only maximal safe paths with positive excess flow are the following:
$(\{a, b, c, d, e, f\}, w=3)$, $(\{c, d, e, f, h, i, j\}, w=3)$, 
$(\{a, c, d, e, f, g, h, i, j\}, w=3)$, 
$(\{c, d, e, f, g, h, i, j, l\}, w=3)$, 
$(\{h, i, j, k, l\}, w=3)$. 
The maximal safe paths of the graph are represented using the following three formats (see \Cref{tab:oldRep}). 

\begin{enumerate}
    \item \textbf{Raw.} The paths are reported \textit{explicitly} as a list of vertices followed by the excess flow. 
    
\item \textbf{Concise.}
A \textit{minimal} (not necessarily minimum) set of paths $\cal P$ such that every safe path is a subpath of some path $P_k\in \cal P$, exploiting the \textit{overlaps} among safe paths. All the safe paths that are subpaths of $P_k$ are stored in $I_k$, storing $(start,end, excess~flow)$.
 This demonstrates that there isn't a unique concise representation, and different concise representations can have different costs. 

\item \textbf{Optimal.} Each safe path is reported using an edge and one/two endpoints using the following property.

\begin{theorem}[Representative edge~\cite{optFlowTheory}]
Every maximal safe path $P$ of the flow decomposition of a DAG has a representative edge $e_P$, where $P$ is $e_P$ with its left and right extensions, where
\begin{enumerate}
    \item Trivial Safe Paths include left extensions of a unique maximum incoming edge.
    \item Non-trivial Safe Paths include left and right extensions of a non-unique maximum incoming edge.
\end{enumerate}
\label{thm:repE}
\end{theorem}

For a unique maximum incoming edge, we thus get a unique maximal (trivial) safe path by using the maximum left extension that is safe. For a non-unique maximum incoming edge, we can get multiple maximal safe paths by considering different combinations of left and right extensions (recall extension lemma). Thus, trivial safe paths are represented using the edge and left node with excess flow, and non-trivial paths are represented using the edge with multiple pairs of left and right nodes with the corresponding excess flows. 
\end{enumerate}

\begin{table}[h]
    \centering
    \begin{tabular}{|l|l|l|l|}
    \hline
       \texttt{Raw} $(41)$  &  \texttt{Concise} $(39)$ & \texttt{Optimal Concise} $(36)$ & \texttt{Optimal} $(23)$ \\
    \hline
       $a~b~c~d~e~f~3$  & $a~b~c~d~e~f~h~i~j$ & $a~c~d~e~f~g~h~i~j~l$ & $b~c$\\
       $c~d~e~f~h~i~j~3$  & $a~f~3$ & $a~j~3$ & $a~f~3$\\
       $a~c~d~e~f~g~h~i~j~3$  & $c~j~3$ & $c~l~3$ & $f~h$\\
      $c~d~e~f~g~h~i~j~l~3$  & $a~c~d~e~f~g~h~i~j~l$ & $a~b~c~d~e~f~h~i~j~k~l$ & $c~j~3$\\
      $h~i~j~k~l~3$  & $a~j~3$ & $a~f~3$ & $k~l$\\
      & $c~l~3$ & $c~j~3$ & $h~l~3$ \\
      & $h~i~j~k~l$ & $h~l~3$ & $a~i~j~3$\\
      & $h~l~3$ & & $c~j~l~3$\\ \hline
    \end{tabular}
    \caption{Safe paths in various representations for the example graph in \Cref{fig:baseGraph}.}
    \label{tab:oldRep}
\end{table}

\label{sec:sota}
Khan et al.~\cite{KhanMMLA22} computed a candidate flow decomposition using the classical approach~\cite{FordNetwork} having $O(m)$ paths each of $O(n)$ length. Since a safe path is also a subpath of the above decomposition, they verified the safety for all its subpaths. They use a two-pointer scan to find the maximal safe prefix for each path in the flow decomposition by extending the first edge to the right using \Cref{lem:excess-flow}. The next maximal safe path is computed by first extending it to the right (making the excess flow negative), then reducing it from the left until it becomes positive again, and then extending it further to the right to make it right maximal. Thus, all the maximal safe subpaths of each path in the flow decomposition can be computed in time proportional to the total length, which is $O(mn)$ in the worst case. 
\begin{enumerate}
    \item \textbf{Raw.} For reporting the safe path, every maximal safe path is printed from the left to the right pointer, which is $O(mn^2)$ in the worst case. 
    \item \textbf{Concise.} The consequent and overlapping safe paths on a path in the flow decomposition are merged, which is reported along with the indices of the maximal safe paths. 
\end{enumerate}

\textbf{Limitations.}  Since the flow decomposition may have several overlapping subpaths, the above approach (for both raw and concise) computes duplicate paths as well as prefixes and suffixes because a maximal safe subpath of a path may not be the maximal safe path in the graph. These duplicates, prefixes, and suffixes are later removed using an AC Trie. Since the intermediate output $out_{tmpR}$ can be much larger than the true size of raw output $out_r$, the actual time complexity of the approach is $O(mn+|out_{tmpR}|+|out_r|)$. Note that $|out_{tmpR}|$ can be $O(mn^2)$ despite $|out_r|$ being significantly smaller. Similarly for the concise output $out_c$ the actual time complexity is $O(mn+|out_{tmpC}|+|out_c|)$. Note that $|out_{tmpC}|$ can be $O(mn)$ despite $out_c$ being significantly smaller. Further, note that the concise output produced is uniquely dependent on the flow decomposition used. Hence, they do not guarantee the optimality of the concise output.

Khan and Tomescu~\cite{optFlowTheory} addressed these limitations by processing the graph in topological order to avoid reporting duplicate and non-maximal paths. They maintain all left maximal safe paths ending at the currently processed vertex and build a trie-like data structure for paths merging common suffixes. All the paths are extended to the maximum outgoing neighbour if their excess flows remain positive. The remaining out neighbours compute their left extensions along the maximum incoming edges until the excess flow becomes negative. Thus, the entire graph is processed once in $O(m)$ time.
\begin{enumerate}
    \item \textbf{Raw.} The maximal safe paths are reported whenever its right extension becomes unsafe on extending along the unique maximum outgoing edge, requiring $O(m+|out_r|)$ time.
    \item \textbf{Concise.} The maximal safe paths ending at a vertex are appended to the new safe paths created with the unique maximum outgoing edge. The other paths generated using non-unique maximum outgoing edges are distributed among the paths whose safe paths end at the current vertex, producing an optimal concise representation $out^*_c$ in $O(m+|out^*_c|)$ time.
    \item \textbf{Optimal.} Using the concept of \textit{representative edges} (\Cref{thm:repE}), both trivial paths and non-trivial paths can be computed by performing a binary search on the forests generated by the unique maximum incoming and outgoing edges. Some additional checks allow the binary searches to be limited to the cases when a maximal path is reported, requiring overall $O(m+|out_o|\log n)$ time, where $out_o$ is the output in the optimal representation.
\end{enumerate}

\textbf{Implementations Issues.} Despite the clear description of the above algorithms, during the practical implementation of the same, we come across some challenges. The pseudocode of the algorithm for optimal concise representation had several typos, resulting in logical mistakes. The corrected pseudocode of the same is presented in~\Cref{second-algo}. Further, in the case of optimal representation, the pseudocode was not provided, leading to significant additional checks and adjustments required to implement the desired intuition. In particular, to avoid creating duplicates, suffixes, and prefixes and to avoid the binary search if no maximal safe path is present, some additional details were required. The comprehensive algorithm and its pseudocode are presented in~\Cref{Third-algo-code}.

\section{Experimental Evaluation}\label{sec:experiment-evaluation}
We now evaluate the performance of all the algorithms on real and random graphs.

\subsection{Algorithms and Implementation details}\label{section:base-algos}
We evaluate the following algorithms:
\begin{enumerate}
    \item \texttt{RawRep.} The previous algorithm~\cite{KhanMMLA22} for raw representation.
    \item \texttt{ConRep.} The previous algorithm~\cite{KhanMMLA22} for concise representation.
    \item \texttt{OptRawRep.} The optimal algorithm~\cite{optFlowTheory} for raw representation.
    \item \texttt{OptConRep.} The modified  algorithm~\cite{optFlowTheory} for concise representation (see \Cref{second-algo}).
    \item \texttt{OptRep.} The modified algorithm~\cite{optFlowTheory} for optimal representation (see~\Cref{Third-algo-code}).
\end{enumerate}

Khan et al.~\cite{KhanMMLA22} reported the time and memory of their algorithms without using AC Trie to remove duplicates, suffixes, and prefixes, leading to an unfair comparison with the new algorithms. Thus, our evaluation (\texttt{RawRep} and \texttt{ConRep}) includes the time and memory required by AC Trie (see performance measures of original algorithms in~\cref{apn:extraTables}). Further, their implementation~\cite{KhanMMLA22} stored the entire input in a string before printing it, which seemed wasteful, impacting the memory requirements of the algorithm. Our implementations print the output directly after computation without storing them \textit{explicitly}.



The optimal representation algorithm uses level ancestor data structure~\cite{mabrey2021static} theoretically requiring $O(1)$ time per query. However, in practice~\cite{mabrey2021static}, a simpler algorithm~\cite{LevelAncestor} performs better despite having $O(\log n)$ worst-case bound. We thus used the latter algorithm, as was previously done in an experimental study of Incremental DFS algorithms~\cite{BaswanaG018}.



\subsection{Evaluation Metrics}
Our algorithms are primarily evaluated on the performance measures of \textit{Execution Time} (time) and  \textit{Memory Usage} (memory). In addition, we compare to \textit{Output Size} (space) for each representation. Khan et al.~\cite{KhanMMLA22} additionally compared the quality of the solution using metrics such as \textit{Weighted precision, Maximum relative coverage, F-score}. However, for all the evaluated algorithms (safe and complete), these metrics would remain exactly the same and hence can be skipped in our evaluation.

\subsection{Environment Details}


All the evaluated algorithms are implemented in C++, while the Python scripts were used to compute the metrics. The algorithms were compiled using GNU C++ compiler version 11.1.0, with \texttt{-O3} optimization flag.  
The performance of the algorithms was evaluated on the PARAM Ganga\footnote{https://hpc.iitr.ac.in} supercomputer system at IIT Roorkee, using a single core of Intel Xeon Platinum 8268 CPU 2.90GHZ with 16GB memory 
The source code of our project is available on GitHub\footnote{https://github.com/Bashar-Ahmed/Safe-Flow-Decomposition} under GNU General Public License v3 license.


\section{
Evaluation on Real Datasets}\label{sec:evaluation-Real}
We focus on the RNA Assembly problem and consider the flow graphs extracted as the splice graphs of simulated RNA-Seq experiments. Given a set of RNA transcripts, their expression levels are simulated, and the transcripts are superimposed to generate a flow graph. Further, we restrict our study to the \textit{perfect} scenario to highlight the significance of each approach, removing the biases introduced by real RNA-Seq experiments. Thus, we consider the datasets used in the previous experimental study on safe flow decomposition~\cite{KhanMMLA22}.


\subsection{Datasets}\label{sec:datasets}

We consider the two RNA transcripts datasets generated based on the approach of Shao et al.~\cite{Shao}, which were used by Khan et al.~\cite{KhanMMLA22}. As in the case of~\cite{KhanMMLA22}, we skip the funnel instances to avoid skewing our results. Recall (\Cref{sec:prelims}) that the safety of funnels is trivial as they have a unique flow decomposition. 


\textit{Catfish dataset:} We examine the dataset initially utilised by Shao and Kingsford~\cite{Shao}, comprising 100 simulated human transcriptomes created with~\cite{Flux-simulator} for humans, mouse, and zebrafish. It also contains 1000 experiments from the Sequence Read Archive, where transcript abundances were simulated using Salmon~\cite{Salmon}. Additionally, it also includes 30 simulated random experiment instances, which were added later to this dataset. The simulation instance contains 3000 graphs, all of which are non-trivial and are relatively larger graphs ranging from vertex count 100 to 1000, which makes it a better dataset to help amplify the performance difference between our algorithms and the existing ones. Several benchmarking studies~\cite{BenchmarkingStudies1,BenchmarkingStudies2} on flow decomposition have also used this dataset. There is a total of 17,338,407 graphs in this dataset, including 8,304,682 (47.90\%) non-funnel graphs. For the sake of evaluation, we distinguish the various sources from which the underlying graph is derived.  \par
\textit{Reference-Sim dataset:} We consider a dataset~\cite{Ref-Sim} containing a single simulated transcriptome. There is a total of 17,941 graphs in this dataset, of which 16,651 (92.8\%) graphs have at least one node, including 10,323 (57.54$\%$) non-funnel graphs.

\begin{table}[!h]
    \centering
    \begin{tabular}{|P{2.2cm}|P{1.5cm}|P{1.2cm}|P{1.2cm}|P{1.8cm}|P{1.8cm}|}
    \hline
Dataset & Number of Graphs & Average Nodes & Average Edges & Average Complexity & Funnel Probability\\ 
\hline
Zebrafish & 413,470 & 18.78 & 21.29 & 2.34 & 0.94 \\
\hline
Mouse & 440,005 & 19.06 & 22.73 & 2.80 & 0.91 \\
\hline
Human & 502,390 & 19.07 & 23.10 & 2.87 & 0.88 \\
\hline
Salmon & 6,945,817 & 20.79 & 26.95 & 3.79 & 0.82 \\
\hline
Ref-Sim & 10,323 & 30.15 & 45.72 & 8.15 & 0.71 \\
\hline
Simulation & 3,000 & 841.33 & 2659.35 & 103.33 & 0.23 \\
\hline
    \end{tabular}
    \caption{Description of the Datasets 
    }\label{tab:results-all}
\end{table}

Table~\ref{tab:results-all} contains the description of the catfish and reference-sim datasets ordered by the average number of nodes per graph. As evident,
the reference-sim dataset (Ref-Sim) has the average number of nodes between the salmon and simulation datasets from Catfish. As evident from the table, most parameters of the datasets in the above order follow a similar behaviour. The complexity of a flow graph is the number of paths in its true path decomposition. In addition, we mention the funnel probability for a ratio of vertices having funnel-like property (unit indegree or outdegree), which is useful for studying random graphs.

\subsection{Results}\label{sec:results}
The relative performance measures of all the evaluated algorithms are described in \Cref{tab:relPerf}. \texttt{OptRawRep} requires the minimum time for most datasets followed by \texttt{OptRep}, while \texttt{ConRep} and \texttt{RawRep} perform the worst. \texttt{OptRawRep} improves \texttt{RawRep} by $4-60\%$ in time and $5-85\%$ in memory, while \texttt{OptConRep} improve \texttt{ConRep} by $6-70\%$ in time and  $8-76\%$ in memory, where improvement increases with the size of the graph. This general trend of newer algorithms performing better than the older ones can be attributed to the $O(mn)$ extra factor to compute and process a candidate flow decomposition. 

However, the dependence on the \textit{output size} of the newer algorithms results in inverse expected performance, eg. \texttt{OptRawRep} performs better than \texttt{OptConRep} and \texttt{OptRep} despite having larger expected output size. This can be attributed to a \textit{potentially} smaller reduction in the output size as compared to the added complexity of the algorithm, requiring further investigation. The only exception is Salmon and Ref-Sim, where \texttt{OptRep} performs equivalent and better than \texttt{OptRawRep}, respectively. However, this change does not carry on to the Simulation dataset where the relative performance of \texttt{OptRep} is even worse. The memory requirements also show a similar trend with the exception of Salmon, where the least memory is required by \texttt{OptConRep}. The older algorithms require higher relative memory as the size of the graph increases, which is again evident from storing the candidate flow decomposition. 

\begin{table}[!h]
\centering
\begin{tabular}{|c|c|c|c|c|c|c|c|} \hline
Algorithm & Parameter & Zebrafish & Mouse & Human & Salmon & Ref-Sim & Simulation \\ \hline
\multirow{2}{*}{\texttt{RawRep}} & Time & 1.04x & 1.09x & 1.07x & 1.16x & 1.33x & 1.59x \\
 & Memory & 1.05x & 1.12x & 1.03x & 1.07x & 1.26x & 1.85x \\ \hline
\multirow{2}{*}{\texttt{ConRep}} & Time & 1.1x & 1.21x & 1.2x & 1.37x & 2.04x & 1.74x \\
 & Memory & 1.03x & 1.13x & 1.02x & 1.09x & 1.69x & 1.89x \\ \hline
\multirow{2}{*}{\texttt{OptRawRep}} & Time & \textbf{1.0x} & \textbf{1.0x} & \textbf{1.0x} & \textbf{1.0x} & 1.04x & \textbf{1.0x} \\
 & Memory & \textbf{1.0x} & \textbf{1.0x} & \textbf{1.0x} & 1.01x & \textbf{1.0x} & \textbf{1.0x} \\ \hline
\multirow{2}{*}{\texttt{OptConRep}} & Time & 1.04x & 1.08x & 1.09x & 1.09x & 1.19x & 1.22x \\
 & Memory & 1.01x & 1.01x & 1.01x & \textbf{1.0x} & 1.01x & 1.07x \\ \hline
\multirow{2}{*}{\texttt{OptRep}} & Time & 1.04x & 1.04x & 1.05x & \textbf{1.0x} & \textbf{1.0x} & 1.21x \\
 & Memory & 1.04x & 1.04x & 1.03x & 1.04x & 1.13x & 1.12x \\ \hline
    \end{tabular} 
    \caption{Relative Performance (Actual Performance in \Cref{tab:actP}, \Cref{apn:extraTables}) of algorithms.}\label{tab:relPerf}
\end{table}

\begin{table}[!h]
\centering
\begin{tabular}{|c|c|c|c|c|c|c|} \hline
Algorithm & Zebrafish & Mouse & Human & Salmon & Ref-Sim & Simulation \\ \hline
\texttt{RawRep} & 2.35x & 2.08x & 1.9x & 1.88x & 1.7x & \textbf{1.0x} \\ \hline
\texttt{ConRep} & 2.7x & 2.94x & 2.73x & 3.09x & 4.92x & 1.53x \\ \hline
\texttt{OptRawRep} & 2.35x & 2.08x & 1.9x & 1.88x & 1.7x & \textbf{1.0x} \\ \hline
\texttt{OptConRep} & 2.21x & 2.03x & 1.9x & 1.76x & 1.81x & 1.3x \\ \hline
\texttt{OptRep} & \textbf{1.0x} & \textbf{1.0x} & \textbf{1.0x} & \textbf{1.0x} & \textbf{1.0x} & 1.08x \\ \hline
    \end{tabular} 
    \caption{Relative Output Size (Actual Output Size in \Cref{tab:actS} in \Cref{apn:extraTables}) of algorithms.}\label{tab:relSize}
\end{table}

We thus investigate the relative output size of the various algorithms (and hence the representations) in \Cref{tab:relSize}. As expected, the raw output generated by \texttt{RawRep} and \texttt{OptRawRep} matches exactly. On the other hand, the concise output generated by \texttt{OptConRep} improves \texttt{ConRep} by $17-170\%$, which increases with the size of the graph (with the exception of Simulation). Further, as expected, the \texttt{OptRep} requires the minimum space improving other representations by $70-135\%$, except for Simulation, where the raw representation is the best. Also, \texttt{ConRep} is, in general, worse than \texttt{RawRep} while \texttt{OptConRep} improves it except in the case of Ref-Sim and Simulation.

\begin{figure}[H]
\begin{subfigure}[t]{.34\textwidth}
\includegraphics[trim=5 7 10 7, clip,scale=.14]{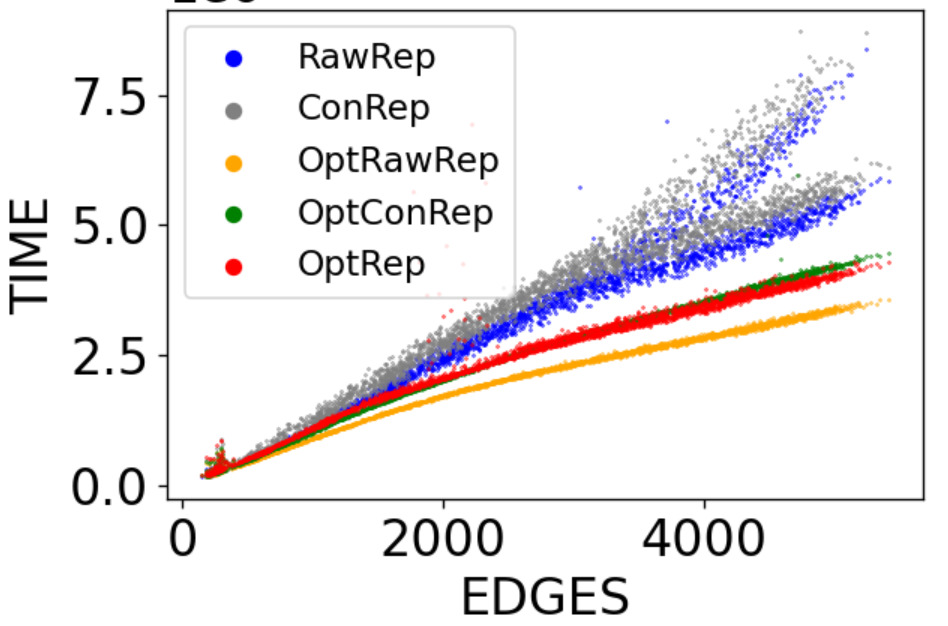}
    \caption{Time ($sec$)}
    \label{fig:timeVedge}
\end{subfigure}
\begin{subfigure}[t]{.32\textwidth}
    \includegraphics[trim=5 7 10 7, clip,scale=.14]{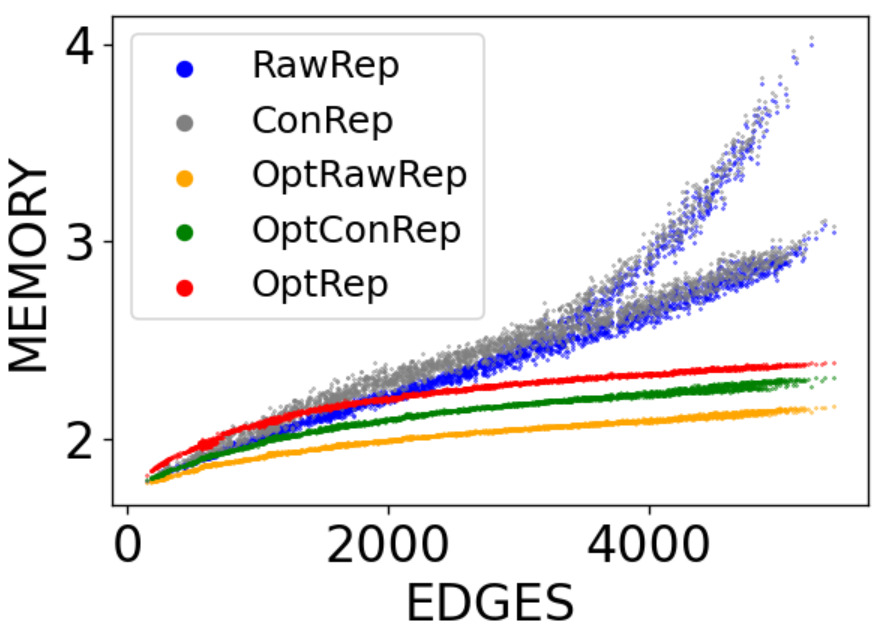}
    \caption{Memory ($MB$)}
    \label{fig:memVedge}
\end{subfigure}
\begin{subfigure}[t]{.32\textwidth}
    \includegraphics[trim=5 7 10 7, clip,scale=.14]{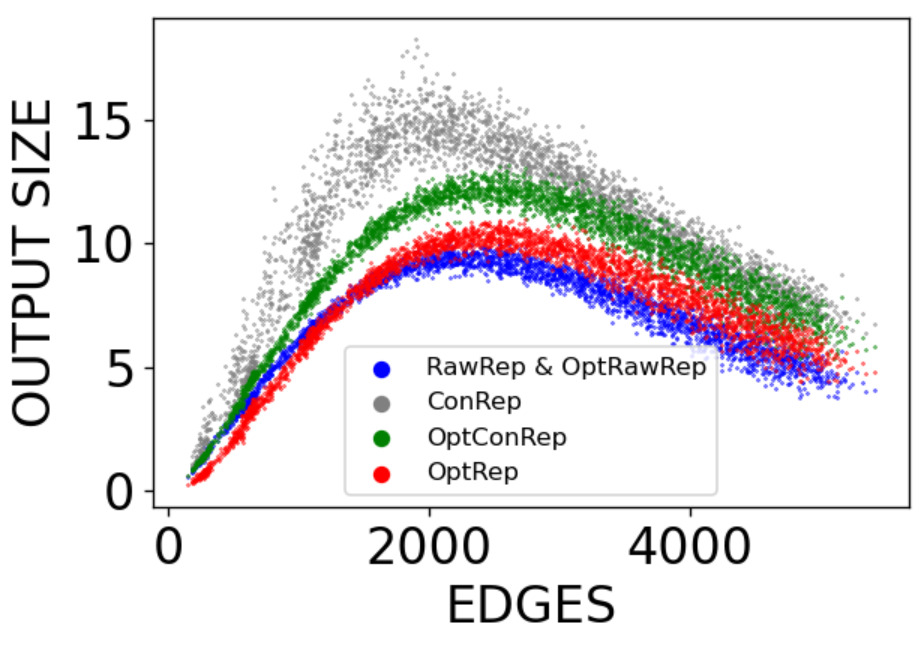}
    \caption{Output Size ($MB$)}
    \label{fig:sizeVedge}
\end{subfigure}
\caption{Performance measures of the algorithms for Simulation with respect to number of edges.}
\label{fig:plotsVedgesSim}
\end{figure}

Finally, we plot in \Cref{fig:plotsVedgesSim} the performance of the algorithms with respect to the corresponding size of the graph (edges), as the complexity of most algorithms is directly proportional to the number of edges along with the output size. For simplicity, we only show the dataset having significantly larger graphs, i.e., the Simulation dataset. 
Similar plots for increasing nodes, as well as the plots for the other datasets, are added to \Cref{apn:extraPlots}. 

\begin{figure}[h]
\begin{subfigure}[t]{.32\textwidth}
    \includegraphics[trim=5 7 10 7, clip,scale=.14]{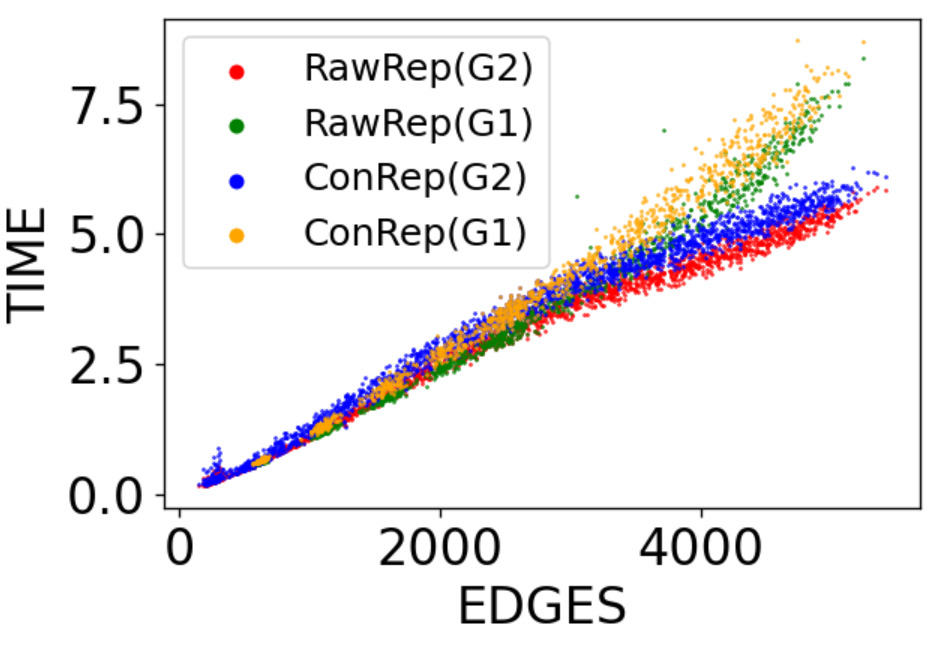}
    \caption{Time ($sec$) for G1 and G2}
    \label{fig:timevsedges-split}
\end{subfigure}\hfill
    \begin{subfigure}[t]{.32\textwidth}
    \includegraphics[trim=5 7 10 7, clip,scale=.14]{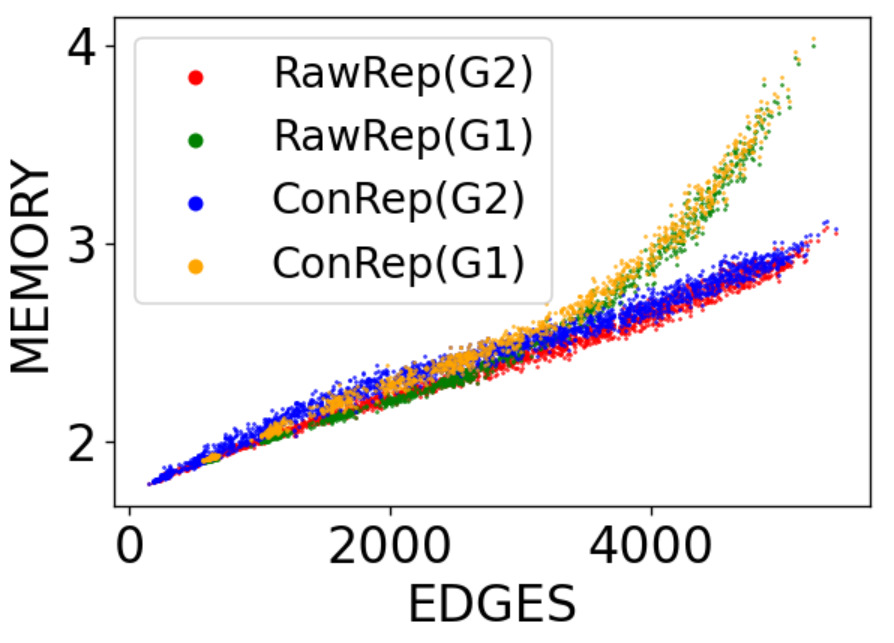}
    \caption{Memory ($MB$) for G1 and G2}
\label{fig:memoryvsedges-split}
\end{subfigure}
\begin{subfigure}[t]{.32\textwidth}
    \includegraphics[trim=5 7 10 7, clip,scale=.3]{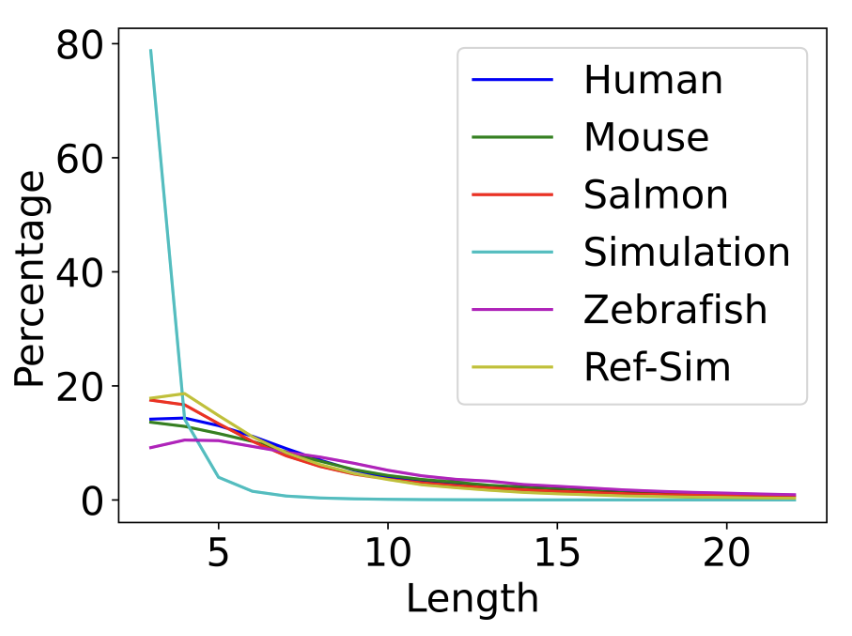}
    \caption{Safe paths of different lengths}
    \label{fig:percentage-safe-paths}
\end{subfigure}
\label{fig:furtherExp}
\caption{Further investigation for understanding observations in results.}
\end{figure}

While the newer algorithms are following an almost uniform trend, the older algorithms show a surprising split in the trend for both time and memory. This motivated us to examine the dataset closely. The Simulation dataset has two types of graphs, \textit{firstly} $G1$ having constant complexity (100), and \textit{secondly} $G2$ having a constant maximum length (50) in the true path decomposition. Apparently, these factors primarily affect the flow decomposition and not the safe paths. Hence, it results in two types of plots for the older algorithms using a candidate flow decomposition. 
For increasing size of the graph, $G1$ thus represents a smaller flow decomposition, as compared to $G2$. 
In the updated graphs (see \Cref{fig:timevsedges-split,fig:memoryvsedges-split}) treating the two types of graphs separately, we see a clear difference in corresponding trends. 

Surprisingly, the output size decreases with the number of edges for all the representations. The decrease in output size of \texttt{OptRep} indicates the decrease in the number of safe paths. Further, as the graph size increases, \texttt{OptRawRep} (or \texttt{RawRep}) becomes smaller than \texttt{OptRep}. Also, as the graph size increases, the benefit of \texttt{OptConRep} over \texttt{ConRep} also decreases. Both these surprises can be explained if the overlap between safe paths and the length of safe paths decreases significantly. 
With the above observations, the time required by the newer algorithms clearly follows the expected trend, increasing with the number of edges and decreasing as the output size decreases. The faster performance of \texttt{OptRawRep} can now be explained using the smaller size of its output. The variation between \texttt{OptConRep} and \texttt{OptRep} can be explained using the complexity of data structures requiring higher time and memory.

Since the behaviour of the simulation dataset was different from the other datasets, we plot the frequency of safe paths of different lengths for all the datasets in \Cref{fig:percentage-safe-paths}. Clearly, the different behaviour of the Simulation dataset can be justified by the percentage of shorter safe paths, which reduces the benefit of concise and optimal representation over raw representation. Consider two safe path of length 2 $a,b,c$ and $b,c,d$. The raw representation stores $6$ nodes and corresponding two flow values. The concise stores a single path $a,b,c,d$ two safe path indices as $a,c$ and $b,d$, i.e. 8 nodes with two flow values. The optimal representation (assuming first trivial and second non-trivial) stores (a) trivial: the edge $b,c$ with two end nodes $a,c$, and (b) non-trivial: the edge $c,d$ and end points $b,d$, i.e., 8 nodes with the two flow values. Hence, for the shorter safe paths, the variation among representations becomes unclear.

\begin{observation}
    We highlight the following observations from the evaluation above.
    \begin{enumerate}
        \item \texttt{OptRawRep} improves time up to $60\%$ and memory up to $85\%$ as graph size increases.
        \item \texttt{OptConRep} improves time up to $70\%$, memory up to $76\%$, and output size up to $170\%$  as graph size increases.
        \item \texttt{OptRep} improves the space required by other representations by up to $135\%$.
    \end{enumerate}
\end{observation}

\noindent
\textbf{Note.} High frequency of very short safe paths severely affect the relative sizes of \texttt{OptConRep}, \texttt{OptRep} favouring \texttt{OptRawRep} resulting in performance inverse of the theoretical expectation.


\section{Proposed Heuristics}\label{sec:heuristics}

Inspired by the observations above and the variation of the number of safe paths as the length increases (\Cref{fig:percentage-safe-paths}), we investigate the datasets on the parameters adversely affecting the size of concise and optimal representations as compared to the raw representation in \Cref{tab:conciseParam}.

For concise representation, the percentage of paths representing a single safe path (\% Single) ranges from $\approx 40-57\%$, where additionally storing the endpoints clearly wastes memory. The percentage of safe paths having either start or end points matching the path (\% Start/end) is $\approx 33-45\%$, where clearly an endpoint is wastefully reported. The percentage of successive safe indices of a path where the end of the current matches the start of the next (\% Successive) is $\approx 0.85-1\%$, where at least one endpoint is wasteful. Additionally, we see that the average number of safe indices for a concise path (Avg. Indices) is $\approx 1.3-1.6$, indicating a lot of concise paths reporting exactly one safe index, which again reports both endpoints wastefully. Finally, the average length of the concise path (Avg. C.Len.) is $5-12$ as compared to the average length of Safe paths (Avg. S.Len) of $\approx 3.5-10$, which shows only a minor overlap captured by the concise representation.

For the optimal representation, the percentage of non-trivial paths (\% Non-trivial) ranges from $\approx 69-76\%$. For trivial paths clearly, storing the right end of the safe path or the left end of the edge is wasteful. Moreover, the average number of non-trivial safe path indices per representative edge (Avg. Indices) is $\approx 1$, showing that in most cases, multiple safe paths do not share the same representative edge. 
%


\begin{table}
\centering
\resizebox{\textwidth}{!}{%
\begin{tabular}{|c|c|c|c|c|c|c|c|c|} \hline
\multirow{2}{*}{Dataset} & Raw & \multicolumn{5}{c|}{Concise} & \multicolumn{2}{c|}{Opt} \\ 
 \cline{2-9}
&Avg. S.Len&  \% Single & \% Start/end & \% Successive & Avg. Indices & Avg. C.Len.&\% Non-trivial & Avg. Indices \\ \hline
Zebrafish& 9.90 & 56.71 & 33.66 & 0.85 & 1.36 & 12.74 & 72.17 & 1.014\\ \hline 
Mouse & 8.81 & 51.2 & 38.13 & 0.98 & 1.42 & 11.63& 75.18 &  1.025\\ \hline 
Human & 8.14& 45.56 & 41.83 & 1.16 & 1.5 & 11.17 & 75.67  & 1.026\\ \hline 
Salmon & 8.01& 39.55 & 44.87 & 0.96 & 1.61 & 10.15 & 74.95 &  1.051\\ \hline 
Ref-Sim & 6.52& 52.5 & 37.26 & 0.86 & 1.41 & 9.03 & 89.03  & 1.042\\ \hline 
Simulation &3.64 & 52.42 & 35.19 & 0.16 & 1.43 & 5.14 & 69.64  & 1.003\\ \hline 
   \end{tabular} }
    \caption{Statistics for various representations}\label{tab:conciseParam}
\end{table}

These properties of the representations of the safe path motivate us to propose some heuristic optimizations in the output format to improve the performance of the concise and optimal representations in terms of both size and time required by their algorithms.

\subsection{Heuristics}
We propose the following heuristics to improve the representations (see \Cref{tab:newRep}).

\begin{table}
    \centering
    \begin{tabular}{|l|l|l|l|}
    \hline
       \texttt{OptConRep} $(36)$  &  \texttt{OptConRep$^\#$} $(32)$ & \texttt{OptRep} $(23)$ & \texttt{OptRep$^\#$} $(20)$ \\
    \hline
       $a~c~d~e~f~g~h~i~j~l$& $a~c~d~e~f~g~h~i~j~l$ & $b~c$ & $b~c$\\
       $a~j~3$ & $j~3$ & $a~f~3$ & $a~f~3$\\
       $c~l~3$ & $c~3$ & $f~h$ & $f~h$\\
     $a~b~c~d~e~f~h~i~j~k~l$ & $a~b~c~d~e~f~h~i~j~k~l$ & $c~j~3$ & $c~j~3$\\
      $a~f~3$ &  $f~3$ & $k~l$ & $k~l$\\
       $c~j~3$ & $c~j~3$ & $h~l~3$  & $h~3$ \\
      $h~l~3$ & $h~3$ & $a~i~j~3$  & $3~a~j$ \\
       & & $c~j~l~3$& $3~c~l$\\ \hline
    \end{tabular}
    \caption{Safe paths in new representations for example graph in \Cref{fig:baseGraph}}
    \label{tab:newRep}
\end{table}


\texttt{OptConRep$^\#$.} For every path $P_k$, since the starting node of the first safe path and the ending node of the last safe path are always included in the representation, we omit these two in the corresponding safe paths. Hence, if a path represents exactly one safe path, we simply report its flow (similar to raw representation). Also, if the ending point of an interval coincides with the starting point of the next interval, we omit the starting point of the next interval. For example, if the complete path is $a$-$b$-$c$-$d$-$e$-$f$-$g$-$h$-$i$-$j$ and the safe paths are $a-b-c$, $b$-$c$-$d$-$e$, $e$-$f$-$g$ and $f$-$g$-$h$-$i$-$j$, with safe flow values $f_1$, $f_2$, $f_3$, $f_4$; then 
the new representation is [\{$a~b~c~d~e~f~g~h~i~j$, \{($c$ $f_1$)\, ($b$ $e$ $f_2$)\, ($g$ $f_3$)\, ($f$ $f_4$)\}\}] instead of [\{$a~b~c~d~e~f~g~h~i~j$, \{($a~c$ $f_1$)\, ($b~e$ $f_2$)\, ($e~g$ $f_3$)\, ($f~j$ $f_4$)\}\}] reported by the original algorithm. Note that the number of tokens in the new representation is $19$, down from $22$.

\texttt{OptRep$^\#$.} For every trivial path, we use only two tokens for its representation, which include the starting and ending nodes of the safe path, as the rest are the unique incoming neighbours. For a non trivial path, if the starting (or ending) nodes coincides with the left (or right) node of the representative edge, then we omit it in the representation, since it saves the space of one tokens in the output. Note that both nodes will never coincide simultaneously as we do not consider paths of length one in our solution. 

\begin{table}[!h]
\centering
\resizebox{.9\textwidth}{!}{%
\begin{tabular}{|c|c|c|c|c|c|c|c|} \hline
Algorithm & Parameter & Zebrafish & Mouse & Human & Salmon & Ref-Sim & Simulation \\ \hline
\multirow{3}{*}{\texttt{OptRawRep}} & Time & 1.05x & 1.01x & 1.02x & 1.02x & 1.06x & \textbf{1.0x} \\
 & Memory & 1.02x & 1.03x & 1.03x & 1.03x & 1.03x & \textbf{1.0x} \\
 & Output Size & 2.8x & 2.48x & 2.26x & 2.25x & 1.99x & 1.17x \\ \hline
\multirow{3}{*}{\texttt{OptConRep}} & Time & 1.08x & 1.09x & 1.11x & 1.11x & 1.21x & 1.22x \\
 & Memory & 1.03x & 1.03x & 1.03x & 1.02x & 1.03x & 1.07x \\
 & Output Size & 2.64x & 2.41x & 2.27x & 2.11x & 2.12x & 1.52x \\ \hline
\multirow{3}{*}{\texttt{OptConRep$^\#$}} & Time & \textbf{1.0x} & \textbf{1.0x} & \textbf{1.0x} & \textbf{1.0x} & 1.09x & 1.13x \\
 & Memory & \textbf{1.0x} & \textbf{1.0x} & \textbf{1.0x} & \textbf{1.0x} & \textbf{1.0x} & 1.04x \\
 & Output Size & 2.34x & 2.11x & 1.97x & 1.82x & 1.78x & 1.13x \\ \hline
\multirow{3}{*}{\texttt{OptRep}} & Time & 1.09x & 1.06x & 1.07x & 1.02x & 1.02x & 1.21x \\
 & Memory & 1.06x & 1.07x & 1.06x & 1.06x & 1.16x & 1.12x \\
 & Output Size & 1.19x & 1.19x & 1.19x & 1.2x & 1.17x & 1.26x \\ \hline
\multirow{3}{*}{\texttt{OptRep$^\#$}} & Time & 1.08x & 1.04x & 1.06x & \textbf{1.0x} & \textbf{1.0x} & 1.2x \\
 & Memory & 1.06x & 1.07x & 1.05x & 1.06x & 1.16x & 1.12x \\
 & Output Size & \textbf{1.0x} & \textbf{1.0x} & \textbf{1.0x} & \textbf{1.0x} & \textbf{1.0x} & \textbf{1.0x} \\ \hline
    \end{tabular} }
    \caption{Relative Time, Relative Memory and Relative Space of algorithms using heuristics.}\label{tab:result-size}
\end{table}

The relative performance measures of the 
 newer algorithms (including heuristics) are described in \Cref{tab:result-size}. \texttt{OptConRep$^\#$} clearly performs better than \texttt{OptConRep} in all the three performance measures, which is attributed to smaller output size of \texttt{OptConRep$^\#$}. While the reduction in memory is slight, on time ($8-11\%$) and output size ($10-25\%$) it is significant. The decrease in the output size is more common in Simulation and Ref-Sim which have a smaller average length of  safe path. \texttt{OptConRep$^\#$} even performs better than \texttt{OptRawRep} with few exceptions in Memory and Time. \texttt{OptRep$^\#$} performs better ($\approx20\%$) than \texttt{OptRep} in terms of the Output Size, though no significant change is observed in Time and Memory. Also, the Output Size of \texttt{OptRep$^\#$} is always smaller than \texttt{OptRawRep} even in Simulation as now \texttt{OptRep$^\#$} uses minimum possible number of tokens to represent a safe path.

\begin{figure}[b]
\begin{subfigure}[t]{.32\textwidth}
\includegraphics[trim=5 7 10 7, clip,scale=.14]{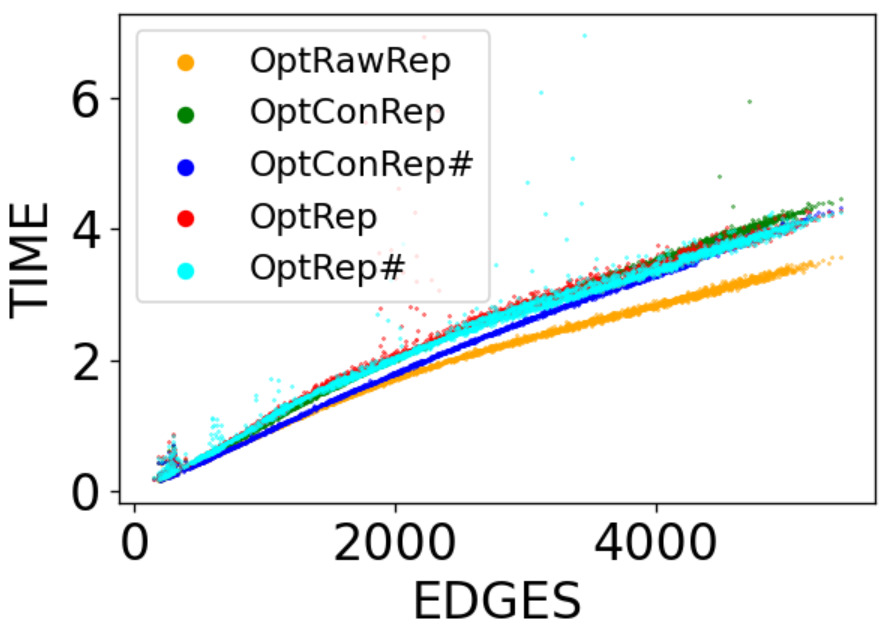}
    \caption{Time (sec)}
    \label{fig:timeVedge-heuristics}
\end{subfigure}
\begin{subfigure}[t]{.33\textwidth}
    \includegraphics[trim=5 7 10 7, clip,scale=.14]{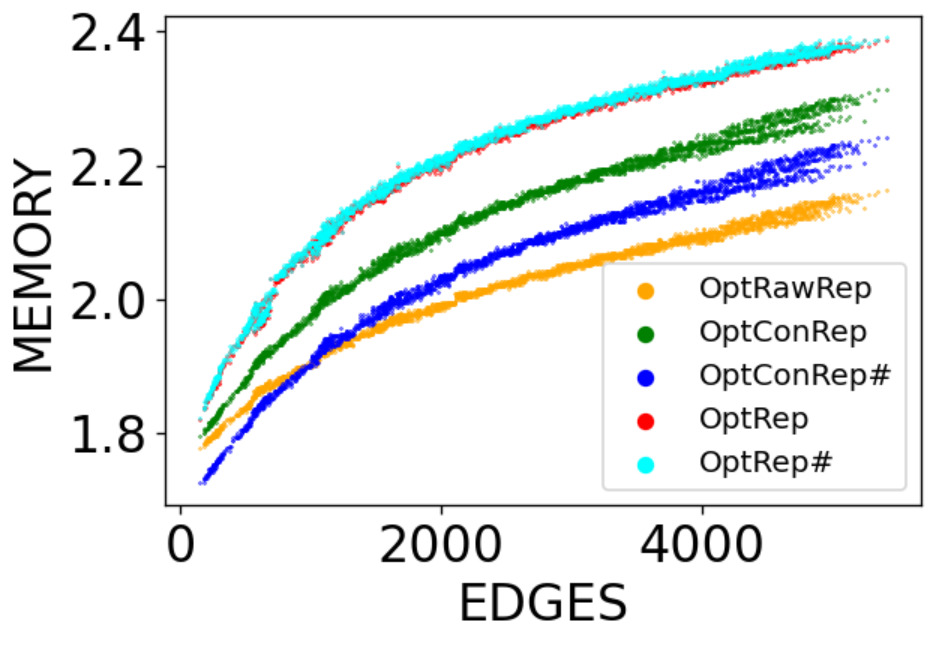}
    \caption{Memory (MB)}
    \label{fig:memVedge-heuristics}
\end{subfigure}
    \begin{subfigure}[t]{.32\textwidth}
    \includegraphics[trim=5 7 10 7, clip,scale=.14]{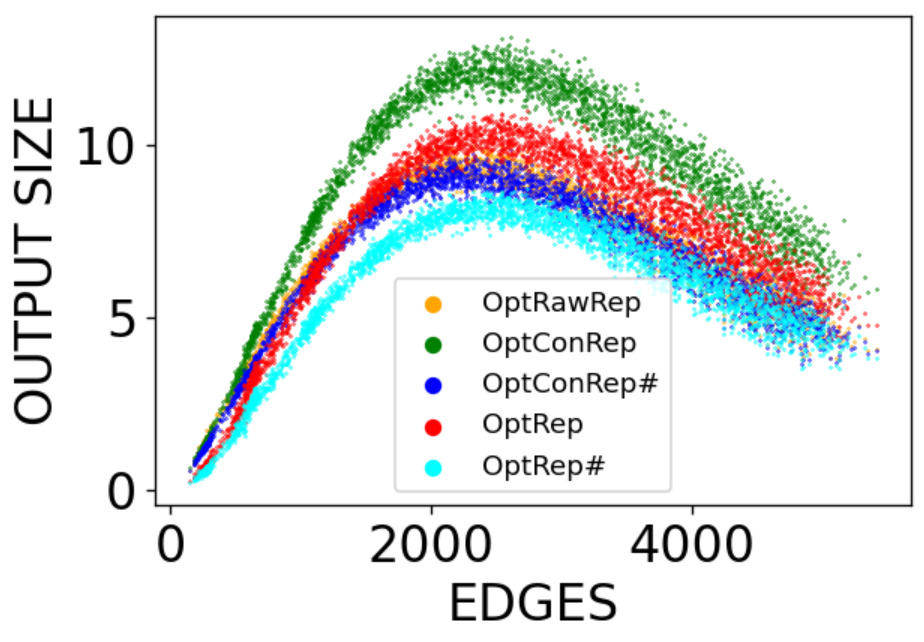}
    \caption{Output Size (MB)}
    \label{fig:sizeVedge-heuristics}
\end{subfigure}
\caption{Performance measures of the algorithms for Simulation with respect to number of edges.}
\label{fig:allVEdgeNew}
\end{figure}

We again plot the performance of the optimal algorithms, along with the heuristics, with respect to the number of edges in \Cref{fig:allVEdgeNew} for Simulation 
(see \Cref{fig:allVnodesNew} in \Cref{apn:extraPlots} for variation with nodes).  
In \Cref{fig:sizeVedge-heuristics}, it is observed that the impact of heuristics on the size increases with the increase in the graph size with \texttt{OptRep} performing the best. The impact on the running time and memory is again larger on concise and negligible on optimal representations. The presence of safe paths of length $2,3$ is thus exploited by the heuristics. 

\begin{observation}
    We highlight the following observations from the evaluation above.
    \begin{enumerate}
        \item \texttt{OptConRep$^\#$} heuristic improves time by $8-10\%$ and output size by $10-25\%$.
        \item \texttt{OptRep$^\#$} heuristic improves output size by around $20\%$. 
    \end{enumerate}
\end{observation}

\section{
Evaluation on Randomly Generated Datasets}\label{sec:random}
The evaluation on real datasets sourced from RNA assembly problem is limited because of the availability of larger datasets. To evaluate the scalability of our results, we further investigate the performance of the algorithms on randomly generated datasets. Further, given the relative performance of old algorithms and new algorithms, we limit our evaluation in this section to the new algorithms with and without heuristics.

\subsection{Datasets}
We consider the following three parameters to generate random graphs: the number of vertices $n$, the number of paths in true decomposition $k$, and the length of each such path $d$. Since we are interested in directed acyclic graphs, we assume the indices of the vertices are according to topological ordering. 

The most commonly studied random graphs are uniformly random graphs (inspired by Erd\H{o}s R\'{e}nyi $G(n, m)$ model), where edges are added uniformly among vertices. However, real data is claimed to be better represented by Power law graphs, where new edges are added biased according to the current degree of the vertices, favoring the vertices with higher degrees. We thus use the following approaches to generating random datasets:

\begin{itemize}
    \item \textbf{Uniformly Random Graphs.} To compute each true path of the decomposition, randomly compute a set of $d-2$ vertices from $[2,n-1]$ and add $d-1$ edges among the consecutive vertices in the topological ordering. We add $k$ such random paths to the graph with a random flow value.
    \item \textbf{Power Law Graphs.} To compute each true path of the decomposition, we compute a set of $d-2$ vertices from $[2,n-1]$ with probability proportional to its degree raised to exponent $3$. Again, we add $d-1$ edges among the consecutive vertices in the topological ordering. We add $k$ such random paths to the graph with a random flow value.
\end{itemize}

\subsection{Results}
The performance of the algorithms is evaluated by varying the three parameters $n,k,d$ one by one, keeping the others constant $n=10000,k=100,d=100$. Hence while evaluating the impact of varying $n$, we keep $k=100,d=100$ and vary $n\in [100,10000]$. Similarly, for the others. 

\begin{figure}[H]
\begin{subfigure}[t]{.34\textwidth}
    \includegraphics[trim=5 7 10 7, clip,scale=.22]{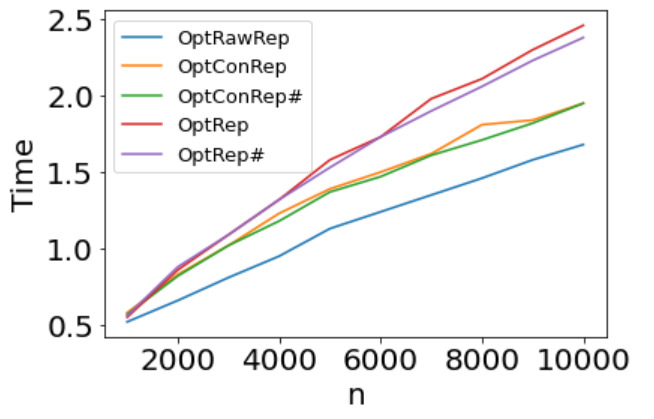}
    \caption{Variation in $n$}
    \label{fig:randomTimeVn}
\end{subfigure}
\begin{subfigure}[t]{.33\textwidth}
\includegraphics[trim=5 7 10 7, clip,scale=.22]{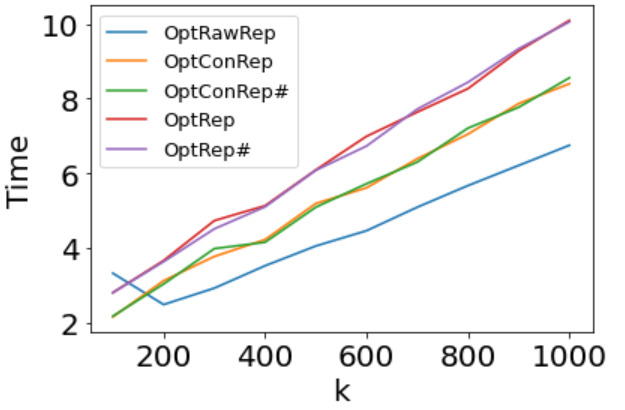}
    \caption{Variation in $k$}
    \label{fig:randomTimeVk}
\end{subfigure}
\begin{subfigure}[t]{.31\textwidth}
\includegraphics[trim=5 7 10 7, clip,scale=.22]{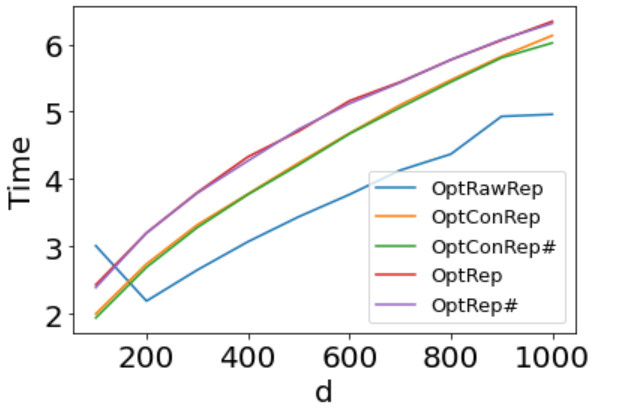}
    \caption{Variation in $d$}
    \label{fig:randomTimeVd}
\end{subfigure}
\caption{Time ($sec$) required by the different algorithms as parameters $n,k,d$ are varied. }
\label{fig:randomTime}
\end{figure}

\begin{figure}[H]
\begin{subfigure}[t]{.33\textwidth}
\includegraphics[trim=5 7 10 7, clip,scale=.22]{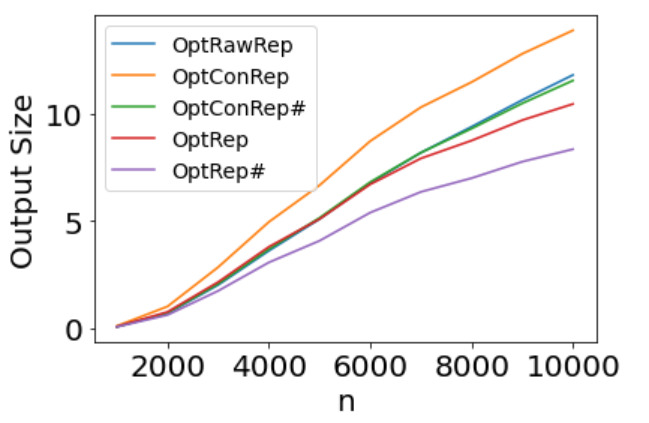}
    \caption{Variation in $n$}
    \label{fig:randomSizeVn}
\end{subfigure}
\begin{subfigure}[t]{.33\textwidth}
    \includegraphics[trim=5 7 10 7, clip,scale=.22]{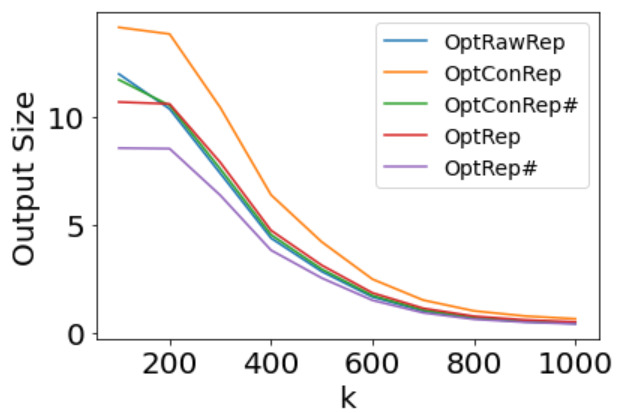}
    \caption{Variation in $k$}
    \label{fig:randomSizeVk}
\end{subfigure}
\begin{subfigure}[t]{.3\textwidth}
    \includegraphics[trim=5 7 10 7, clip,scale=.22]{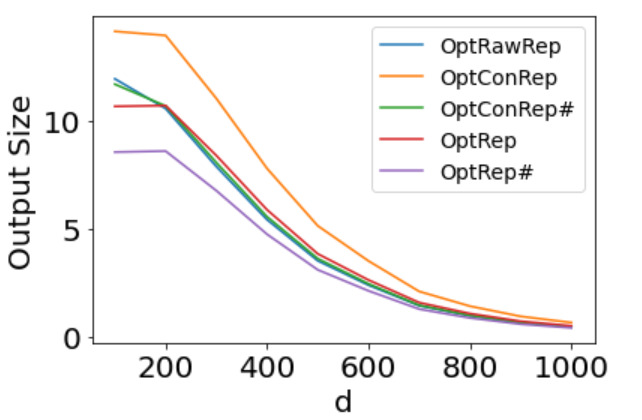}
    \caption{Variation in $d$}
    \label{fig:randomSizeVd}
\end{subfigure}
\caption{Output size ($MB$) of different algorithms as parameters $n,k,d$ are varied.}
\label{fig:randomSize}
\end{figure}

For uniformly random graphs, \Cref{fig:randomTime} shows the time taken by the algorithms in the three experiments. We see that \texttt{OptRawRep} performs the best in all the cases, followed by a nearly equal performance of \texttt{OptConRep, OptConRep$^\#$} followed by the nearly equal performance of \texttt{OptRep, OptRep$^\#$}. To see the relative impact of the output sizes, we also plot the corresponding output sizes in 
\Cref{fig:randomSize}. Notably, the output sizes increase with an increase in $n$ but decrease with the increase of $k,d$. This can be attributed to the potential decrease in the number of safe paths and their lengths in the latter cases.  The impact of heuristics is indeed visible on the output sizes with \texttt{OptConRep$^\#$} nearly matches \texttt{OptRawRep} and \texttt{OptRep}, while \texttt{OptRep$^\#$} has the lowest output size.  Nevertheless, this impact is not reflected in the running times because the output size is insignificant ($0.1\times$) as compared to input size $n$.

\begin{figure}[H]
\begin{subfigure}[t]{.31\textwidth}
    \includegraphics[trim=5 7 10 7, clip,scale=.2]{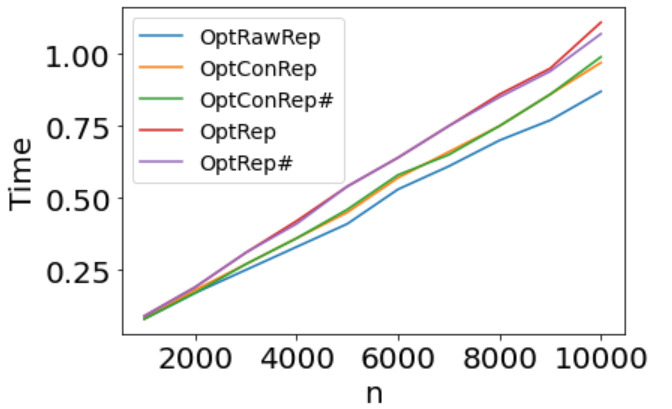}
    \caption{Time taken ($sec$)}
    \label{fig:powerLawTime}
\end{subfigure}
\begin{subfigure}[t]{.31\textwidth}
\includegraphics[trim=5 7 10 7, clip,scale=.2]{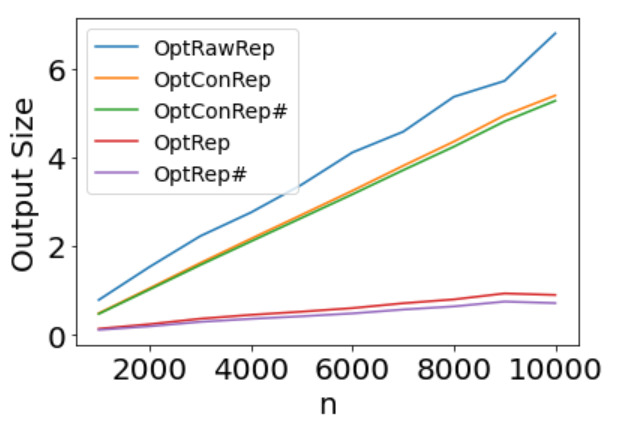}
    \caption{Output size ($MB$)}
    \label{fig:powerLawSize}
\end{subfigure}
\begin{subfigure}[t]{0.34\textwidth}
    \includegraphics[trim=0 0 0 15, clip,scale=.3]{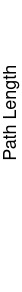}
    \includegraphics[trim=5 7 10 7, clip,scale=.2]{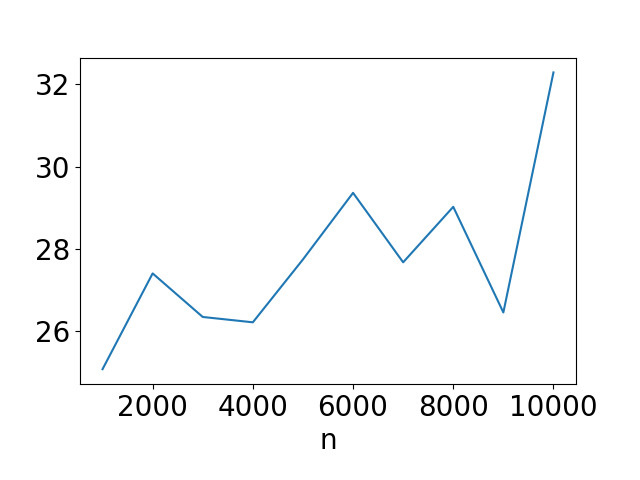}
    \caption{Average safe path length}
    \label{fig:powerLawLength}
\end{subfigure}
\caption{Performance of the algorithms on Power law Graphs as parameter $n$ is varied.}
\label{fig:powerLaw}
\end{figure}

For power law graphs, \Cref{fig:powerLaw} shows the time taken by the algorithms and their corresponding output size only for varying $n$ keeping $k,d$ constant. 
The experiments of the remaining two types (varying $k$ and varying $d$) are omitted here in the interest of space.
Again, \texttt{OptRawRep} performs the best in all the cases, followed by a nearly equal performance of \texttt{OptConRep, OptConRep$^\#$} followed by the nearly equal performance of \texttt{OptRep, OptRep$^\#$}. The impact of the heuristics on output size is relatively insignificant in this case. \texttt{OptRawRep} has the worst output size followed by \texttt{OptConRep,OptConRep$^\#$} and then exceptionally good \texttt{OptRep,OptRep$^\#$}. Again, this impact is not reflected in the relative running times because the total output size is insignificant ($.07\times$) compared to the input size $n$.

\begin{figure}[H] 
\begin{subfigure}[t]{0.32\textwidth}
    \includegraphics[trim=0 0 0 0, clip,scale=.27]{Results/PathLen.jpg}
   \includegraphics[trim=5 7 10 7, clip,scale=.19]{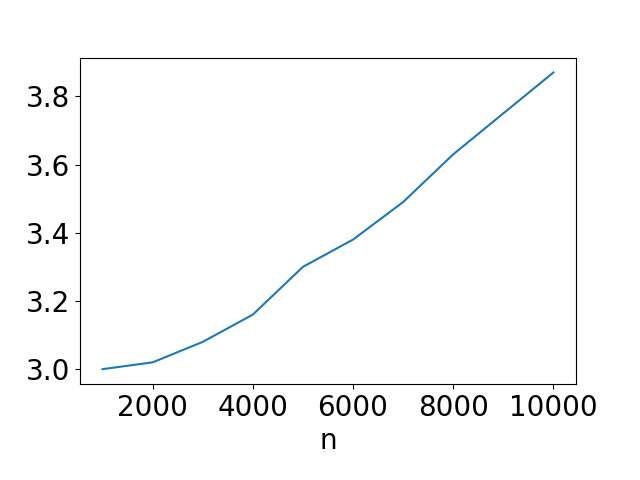}
    \caption{For varying $n$}
    \label{fig:averageLvN}
\end{subfigure}
\begin{subfigure}[t]{0.32\textwidth}
    \includegraphics[trim=0 0 0 0, clip,scale=.27]{Results/PathLen.jpg}
   \includegraphics[trim=5 7 10 7, clip,scale=.19]{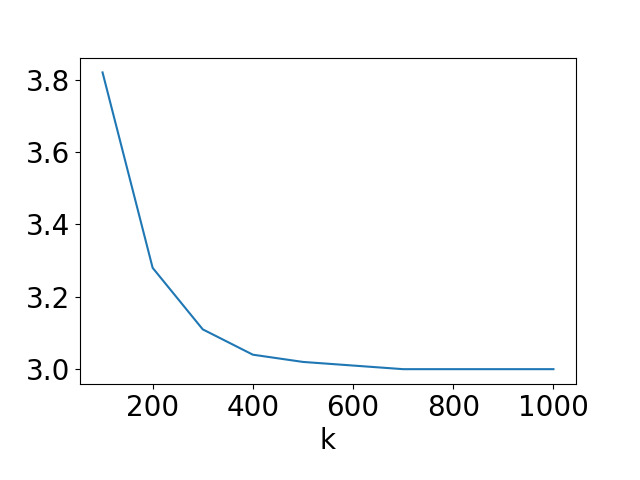}
    \caption{For varying $k$}
    \label{fig:averageLvK}
\end{subfigure}
\begin{subfigure}[t]{0.32\textwidth}
    \includegraphics[trim=0 0 0 0, clip,scale=.27]{Results/PathLen.jpg}
   \includegraphics[trim=5 7 10 7, clip,scale=.19]{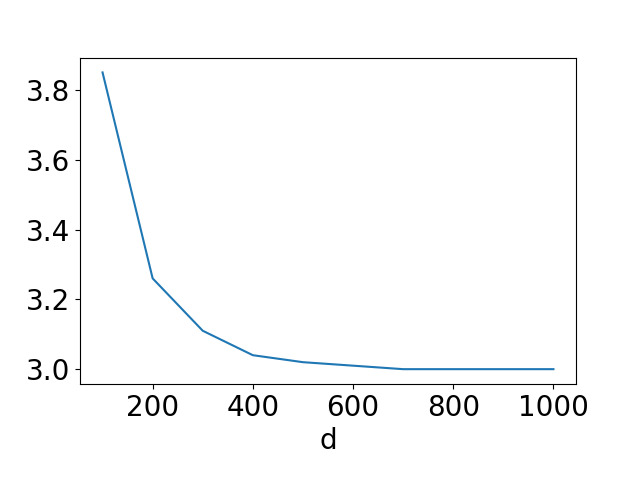}
    \caption{For varying $d$}
    \label{fig:averageLvD}
\end{subfigure}
\caption{Average length of safe paths for uniformly random graphs varying parameters $n,k,d$.}
\label{fig:randomAverageLength}
\end{figure}

On further investigation, we realized the average length of safe paths is $3-3.8$ for uniform random graphs (see \Cref{fig:randomAverageLength}) and around $30$ for power law graphs (see \Cref{fig:powerLawLength}), again insignificant as compared to the input size. Recall that in real graphs, these were of comparable sizes (see \Cref{tab:results-all,tab:conciseParam}). This can be attributed to the probability of a vertex having unit in-degree or out-degree (resulting in funnel-like structure having longer paths). For the evaluated real graphs average funnel probability is $p=81\%$ (\Cref{tab:results-all}). We thus proposed \textit{improved} random graphs more suited to our applications as follows:

\begin{itemize}
    \item \textbf{Improved Random Graphs.} We add a \textit{complete} path of length $n-1$ having all vertices in topological order. Each of $k$ random paths is computed using a random set of $d-2$ vertices from $[2,n-1]$ sorted in topological ordering. For every ordered pair, we either add a direct edge between them or use the complete path between them (probability $p^2$ as both funnel vertices). We add the path to the graph with a random flow value.
\end{itemize}

\begin{figure}[H]
\centering
\begin{subfigure}[t]{.31\textwidth}
   \includegraphics[trim=5 7 10 7, clip,scale=.2]{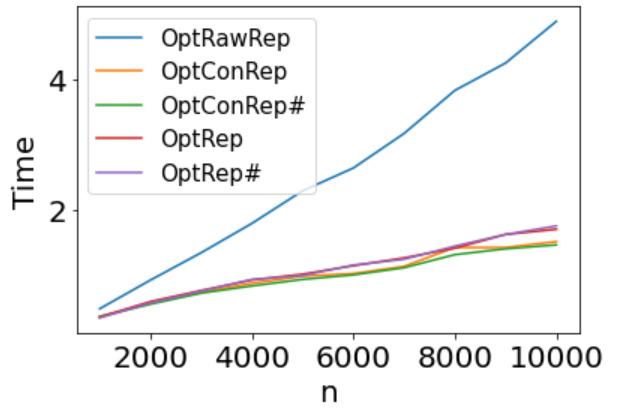}
    \caption{Time taken ($sec$)}
\label{fig:impRandomTime}
\end{subfigure}
\begin{subfigure}[t]{.31\textwidth}
\includegraphics[trim=5 7 10 7, clip,scale=.2]{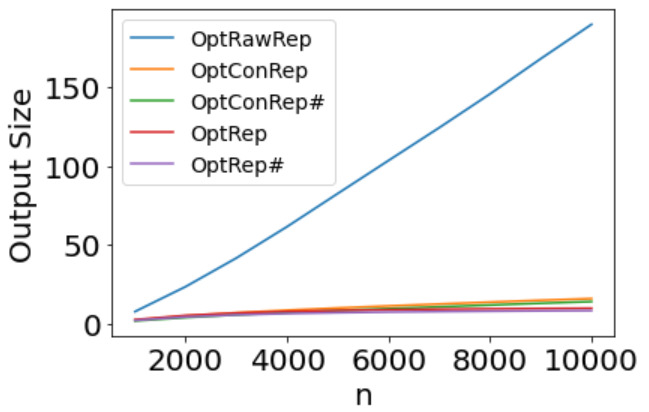}
    \caption{Output size ($MB$)}
\label{fig:impRandomSize}
\end{subfigure}
\begin{subfigure}[t]{0.34\textwidth}
    \includegraphics[trim=0 0 0 15, clip,scale=.27]{Results/PathLen.jpg}
    \includegraphics[trim=5 7 10 7, clip,scale=.2]{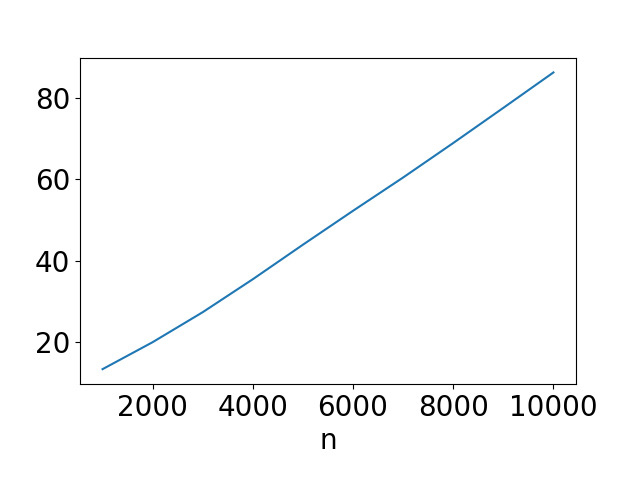}
    \caption{Average safe path length}
\label{fig:impRandomLength}
\end{subfigure}
\caption{Performance of the algorithms on improved random graphs as parameter $n$ is varied.}
\label{fig:impRandom}
\end{figure}

For improved random graphs, \Cref{fig:impRandom} shows the time taken by the algorithms and their corresponding output size only for varying $n$ keeping $k,d$ constant. Again, the experiments of the remaining two types (varying $k$ and varying $d$) are omitted here in the interest of space.  We now see a huge improvement over \texttt{OptRawRep}  for \texttt{OptRep, OptRep$^\#$} closely followed by \texttt{OptConRep,OptConRep$^\#$}. The impact of the heuristics on the output sizes is again relatively insignificant. However, the huge difference between the output size of \texttt{OptRawRep} with the rest explains the performance given that the total output size is comparable to the input size $n$. We show the scalability of our algorithms for some distinct values of $n,k,d$ for improved random graphs (\Cref{tab:snapshots}), which match the performance on real graphs. The new algorithms are indeed scalable, massively improving old ones, which are impractical for large graphs due to the quadratic size of the flow decomposition. \texttt{ConRep} shows output much greater than \texttt{OptRawRep}, probably due to a lot of non-maximal paths added in the concise representation. The improvement in output size is very significant for concise ($92\%$) and optimal ($96\%$) representations over raw representation. The reduction in output size due to heuristics is significant ($14\%$) for concise but insignificant for optimal representation.

\begin{table}[!h]
\centering
\resizebox{\textwidth}{!}{%
\begin{tabular}{|c|c|c|c|c|c|c|c|c|c|c|} \hline
$n$ & $k$ & $d$ & Parameter & \texttt{RawRep} & \texttt{ConRep} & \texttt{OptRawRep} & \texttt{OptConRep} & \texttt{OptConRep}$^\#$ & \texttt{OptRep} & \texttt{OptRep}$^\#$ \\ \hline
\multirow{3}{*}{$1M$} & \multirow{3}{*}{$100$} & \multirow{3}{*}{$10K$} & Time & > $24hr$ & $14.18hr$ & 5.41 & 1.62 & 1.58 & 2.51 & 2.44 \\
 & \multirow{3}{*}{} & \multirow{3}{*}{} & Memory & $312.59GB$ & $379.98GB$ & 310.11 & 336.82 & 338.89 & 730.66 & 730.86 \\
 & \multirow{3}{*}{} & \multirow{3}{*}{} & Output Size & DNF & 7869.74 &  271.13 & 22.01 & 19.21 & 13.18 & 10.94 \\ \hline
 \multirow{3}{*}{$10M$} & \multirow{3}{*}{$100$} & \multirow{3}{*}{$100K$} & Time & > $24hr$ & > $24hr$ & 44.11 & 16.13 & 16.02 & 21.2 & 21.0 \\
 & \multirow{3}{*}{} & \multirow{3}{*}{} & Memory & >$512GB$ & >$512GB$ &  3089.45 & 3377.88 & 3377.09 & 7381.44 & 7381.64 \\
 & \multirow{3}{*}{} & \multirow{3}{*}{} & Output Size & DNF & DNF & 3375.26 & 235.67 & 204.20 & 127.10 & 124.71 \\ \hline
  \multirow{3}{*}{$100M$} & \multirow{3}{*}{$100$} & \multirow{3}{*}{$1M$} & Time & > $24hr$ & > $24hr$ & 441.41 & 162.0 & 162.45 & 221.4 & 223.19 \\
 & \multirow{3}{*}{} & \multirow{3}{*}{} & Memory & >$512GB$ & >$512GB$ & 31103.22 & 33977.62 & 33977.43 & 73063.43 & 73063.43 \\
 & \multirow{3}{*}{} & \multirow{3}{*}{} & Output Size & DNF & DNF & 36618.00 & 2698.87 & 2341.26 & 1361.01 & 1358.59 \\ \hline
\end{tabular}}
\caption{Time ($sec$), Memory ($MB$) and Output Size ($MB$) for improved random graphs having larger values of $n,k$ and $d$ demostrating the scalability of optimal algorithms. DNF:Did Not Finish.}\label{tab:snapshots}
\end{table}

\begin{observation}
    We highlight the following observations from the evaluation above.
    \begin{enumerate}
        \item Random and power law graphs show impact of representations on output size but not time.
        \item Improved random graphs (having comparable input and output size) show massive improvement in time, memory, and output size for newer algorithms and representations. 
        \item The older algorithms are not scalable for large graphs and the representations impact the performance severely.
    \end{enumerate}
\end{observation}

\section{Conclusion}\label{sec:conclusion}
We empirically evaluated the theoretically optimal algorithms and representation for computing safe flow decomposition. These algorithms required additional implementation details for practical evaluation. Our evaluation on real datasets sourced from RNA assembly applications demonstrates significant improvement by optimal algorithms in required time (up to $60-70\%$) and memory (up to $76-85\%$).
Moreover, the optimal concise representation also improves the concise representation (up to $170\%$), and the optimal representation significantly improves all other representations' (up to $135\%$) output size. However, the optimal algorithm for raw representation still proved better due to a large number of very short safe paths. We thus developed heuristics for concise and optimal representations, which significantly impacted the output size ($10-25\%$) and time taken (up to $10\%$). We further evaluated the performance of the algorithms on related random graphs, which showed that only optimal algorithms are scalable for larger graphs ($\geq 1M$ nodes), and follows similar trend as real graphs implying the scalability of our results.

\bibliography{paper}

\begin{thebibliography}{10}

\bibitem{acosta2018safe}
Nidia~Obscura Acosta, Veli M{\"{a}}kinen, and Alexandru~I. Tomescu.
\newblock A safe and complete algorithm for metagenomic assembly.
\newblock {\em Algorithms for Molecular Biology}, 13(1):3:1--3:12, 2018.
\newblock \href {https://doi.org/10.1186/s13015-018-0122-7}
  {\path{doi:10.1186/s13015-018-0122-7}}.

\bibitem{acosta2024simplicity}
Nidia~Obscura Acosta and Alexandru~I Tomescu.
\newblock Simplicity in eulerian circuits: Uniqueness and safety.
\newblock {\em Information Processing Letters}, 183:106421, 2024.

\bibitem{BaaijensRKSS19}
Jasmijn~A. Baaijens, Bastiaan~Van der Roest, Johannes K{\"{o}}ster, Leen
  Stougie, and Alexander Sch{\"{o}}nhuth.
\newblock Full-length de novo viral quasispecies assembly through variation
  graph construction.
\newblock {\em Bioinform.}, 35(24):5086--5094, 2019.
\newblock \href {https://doi.org/10.1093/bioinformatics/btz443}
  {\path{doi:10.1093/bioinformatics/btz443}}.

\bibitem{DBLP:conf/recomb/BaaijensSS20}
Jasmijn~A. Baaijens, Leen Stougie, and Alexander Sch{\"{o}}nhuth.
\newblock Strain-aware assembly of genomes from mixed samples using flow
  variation graphs.
\newblock In {\em Research in Computational Molecular Biology - 24th Annual
  International Conference, {RECOMB} 2020, Padua, Italy, May 10-13, 2020,
  Proceedings}, pages 221--222, 2020.

\bibitem{baier2005k}
Georg Baier, Ekkehard K{\"o}hler, and Martin Skutella.
\newblock The k-splittable flow problem.
\newblock {\em Algorithmica}, 42(3-4):231--248, 2005.
\newblock \href {https://doi.org/10.1007/s00453-005-1167-9}
  {\path{doi:10.1007/s00453-005-1167-9}}.

\bibitem{BaswanaG018}
Surender Baswana, Ayush Goel, and Shahbaz Khan.
\newblock Incremental {DFS} algorithms: a theoretical and experimental study.
\newblock In {\em {SODA}}, pages 53--72. {SIAM}, 2018.

\bibitem{bernard2013flipflop}
Elsa Bernard, Laurent Jacob, Julien Mairal, and Jean{-}Philippe Vert.
\newblock Efficient {RNA} isoform identification and quantification from
  rna-seq data with network flows.
\newblock {\em Bioinform.}, 30(17):2447--2455, 2014.
\newblock \href {https://doi.org/10.1093/bioinformatics/btu317}
  {\path{doi:10.1093/bioinformatics/btu317}}.

\bibitem{cairo2020safety}
Massimo Cairo, Shahbaz Khan, Romeo Rizzi, Sebastian Schmidt, and Alexandru~I.
  Tomescu.
\newblock {Safety in $s$-$t$ Paths, Trails and Walks}.
\newblock {\em Algorithmica}, 84:719--741, 2022.
\newblock URL: \url{https://doi.org/10.1007/s00453-021-00877-w}.

\bibitem{Cairo0RSTZ23}
Massimo Cairo, Shahbaz Khan, Romeo Rizzi, Sebastian~S. Schmidt, Alexandru~I.
  Tomescu, and Elia~C. Zirondelli.
\newblock Cut paths and their remainder structure, with applications.
\newblock In {\em {STACS}}, volume 254 of {\em LIPIcs}, pages 17:1--17:17.
  Schloss Dagstuhl - Leibniz-Zentrum f{\"{u}}r Informatik, 2023.

\bibitem{CairoMART19}
Massimo Cairo, Paul Medvedev, Nidia~Obscura Acosta, Romeo Rizzi, and
  Alexandru~I. Tomescu.
\newblock {An Optimal \emph{O}(\emph{nm}) Algorithm for Enumerating All Walks
  Common to All Closed Edge-covering Walks of a Graph}.
\newblock {\em {ACM} Trans. Algorithms}, 15(4):48:1--48:17, 2019.
\newblock \href {https://doi.org/10.1145/3341731} {\path{doi:10.1145/3341731}}.

\bibitem{cairo2020macrotigs}
Massimo Cairo, Romeo Rizzi, Alexandru~I. Tomescu, and Elia~C. Zirondelli.
\newblock Genome assembly, from practice to theory: Safe, complete and
  linear-time.
\newblock In Nikhil Bansal, Emanuela Merelli, and James Worrell, editors, {\em
  48th International Colloquium on Automata, Languages, and Programming,
  {ICALP} 2021, July 12-16, 2021, Glasgow, Scotland (Virtual Conference)},
  volume 198 of {\em LIPIcs}, pages 43:1--43:18. Schloss Dagstuhl -
  Leibniz-Zentrum f{\"{u}}r Informatik, 2021.

\bibitem{cohen2014effect}
Rami Cohen, Liane Lewin-Eytan, Joseph~Seffi Naor, and Danny Raz.
\newblock On the effect of forwarding table size on sdn network utilization.
\newblock In {\em IEEE INFOCOM 2014-IEEE conference on computer
  communications}, pages 1734--1742. IEEE, 2014.

\bibitem{FordNetwork}
D.~R. Ford and D.~R. Fulkerson.
\newblock {\em Flows in Networks}.
\newblock Princeton University Press, USA, 2010.

\bibitem{gatter2019ryuto}
Thomas Gatter and Peter~F Stadler.
\newblock Ry{\=u}t{\=o}: network-flow based transcriptome reconstruction.
\newblock {\em BMC bioinformatics}, 20(1):190, 2019.
\newblock \href {https://doi.org/10.1186/s12859-019-2786-5}
  {\path{doi:10.1186/s12859-019-2786-5}}.

\bibitem{LevelAncestor}
Dhruv~Matani Gaurav~Menghani.
\newblock A simple solution to the level-ancestor problem.
\newblock 2021.
\newblock \href {https://doi.org/https://doi.org/10.48550/arXiv.1903.01387}
  {\path{doi:https://doi.org/10.48550/arXiv.1903.01387}}.

\bibitem{Flux-simulator}
Thasso Griebel, Benedikt Zacher, Paolo Ribeca, Emanuele Raineri, Vincent
  Lacroix, Roderic Guigó, and Michael Sammeth.
\newblock {Modelling and simulating generic RNA-Seq experiments with the flux
  simulator}.
\newblock {\em Nucleic Acids Research}, 40(20):10073--10083, 09 2012.
\newblock \href
  {http://arxiv.org/abs/https://academic.oup.com/nar/article-pdf/40/20/10073/16958014/gks666.pdf}
  {\path{arXiv:https://academic.oup.com/nar/article-pdf/40/20/10073/16958014/gks666.pdf}},
  \href {https://doi.org/10.1093/nar/gks666} {\path{doi:10.1093/nar/gks666}}.

\bibitem{hartman2012split}
Tzvika Hartman, Avinatan Hassidim, Haim Kaplan, Danny Raz, and Michal Segalov.
\newblock How to split a flow?
\newblock In {\em 2012 Proceedings IEEE INFOCOM}, pages 828--836. IEEE, 2012.

\bibitem{hong2013achieving}
Chi-Yao Hong, Srikanth Kandula, Ratul Mahajan, Ming Zhang, Vijay Gill, Mohan
  Nanduri, and Roger Wattenhofer.
\newblock Achieving high utilization with software-driven wan.
\newblock In {\em Proceedings of the ACM SIGCOMM 2013 conference on SIGCOMM},
  pages 15--26, 2013.

\bibitem{KM95}
John~D. Kececioglu and Eugene~W. Myers.
\newblock Combinatorial algorithms for {DNA} sequence assembly.
\newblock {\em Algorithmica}, 13(1/2):7--51, 1995.

\bibitem{KhanMMLA22}
Shahbaz Khan, Milla Kortelainen, Manuel C{\'{a}}ceres, Lucia Williams, and
  Alexandru~I. Tomescu.
\newblock Safety and completeness in flow decompositions for {RNA} assembly.
\newblock In {\em 26th Annual International Conference, {RECOMB} 2022, San
  Diego, CA, USA, May 22-25, 2022}, pages 177--192, 2022.

\bibitem{optFlowTheory}
Shahbaz Khan and Alexandru~I. Tomescu.
\newblock {Optimizing Safe Flow Decompositions in DAGs}.
\newblock In Shiri Chechik, Gonzalo Navarro, Eva Rotenberg, and Grzegorz
  Herman, editors, {\em 30th Annual European Symposium on Algorithms (ESA
  2022)}, volume 244 of {\em Leibniz International Proceedings in Informatics
  (LIPIcs)}, pages 72:1--72:17, Dagstuhl, Germany, 2022. Schloss Dagstuhl --
  Leibniz-Zentrum f{\"u}r Informatik.
\newblock \href {https://doi.org/10.4230/LIPIcs.ESA.2022.72}
  {\path{doi:10.4230/LIPIcs.ESA.2022.72}}.

\bibitem{BenchmarkingStudies1}
Kyle Kloster, Philipp Kuinke, Michael~P. O'Brien, Felix Reidl,
  Fernando~S{\'{a}}nchez Villaamil, Blair~D. Sullivan, and Andrew van~der Poel.
\newblock A practical fpt algorithm for flow decomposition and transcript
  assembly.
\newblock {\em CoRR}, abs/1706.07851, 2017.
\newblock URL: \url{http://arxiv.org/abs/1706.07851}, \href
  {http://arxiv.org/abs/1706.07851} {\path{arXiv:1706.07851}}.

\bibitem{DBLP:conf/wabi/MaZK20}
Cong Ma, Hongyu Zheng, and Carl Kingsford.
\newblock Exact transcript quantification over splice graphs.
\newblock In {\em 20th International Workshop on Algorithms in Bioinformatics,
  {WABI} 2020, September 7-9, 2020, Pisa, Italy (Virtual Conference)}, pages
  12:1--12:18, 2020.

\bibitem{mabrey2021static}
Matthew Mabrey, Thomas Caputi, Georgios Papamichail, and Dimitris Papamichail.
\newblock Static level ancestors in practice, 2021.
\newblock \href {http://arxiv.org/abs/1402.2741} {\path{arXiv:1402.2741}}.

\bibitem{mumey2015parity}
Brendan Mumey, Samareh Shahmohammadi, Kathryn McManus, and Sean Yaw.
\newblock Parity balancing path flow decomposition and routing.
\newblock In {\em 2015 IEEE Globecom Workshops (GC Wkshps)}, pages 1--6. IEEE,
  2015.

\bibitem{nagarajan2009parametric}
Niranjan Nagarajan and Mihai Pop.
\newblock Parametric complexity of sequence assembly: theory and applications
  to next generation sequencing.
\newblock {\em Journal of computational biology}, 16(7):897--908, 2009.

\bibitem{Salmon}
Rob Patro, Geet Duggal, and Carl Kingsford.
\newblock Salmon: Accurate, versatile and ultrafast quantification from rna-seq
  data using lightweight-alignment.
\newblock {\em bioRxiv}, 2015.
\newblock URL: \url{https://www.biorxiv.org/content/early/2015/06/27/021592},
  \href
  {http://arxiv.org/abs/https://www.biorxiv.org/content/early/2015/06/27/021592.full.pdf}
  {\path{arXiv:https://www.biorxiv.org/content/early/2015/06/27/021592.full.pdf}},
  \href {https://doi.org/10.1101/021592} {\path{doi:10.1101/021592}}.

\bibitem{pertea2015stringtie}
Mihaela Pertea, Geo~M Pertea, Corina~M Antonescu, Tsung-Cheng Chang, Joshua~T
  Mendell, and Steven~L Salzberg.
\newblock Stringtie enables improved reconstruction of a transcriptome from
  rna-seq reads.
\newblock {\em Nature biotechnology}, 33(3):290--295, 2015.
\newblock \href {https://doi.org/10.1038/nbt.3122}
  {\path{doi:10.1038/nbt.3122}}.

\bibitem{PTW01}
Pavel~A. Pevzner, Haixu Tang, and Michael~S. Waterman.
\newblock An {Eulerian} path approach to {DNA} fragment assembly.
\newblock {\em Proceedings of the National Academy of Sciences},
  98(17):9748--9753, 2001.

\bibitem{pienkosz2015integral}
Krzysztof Pie{\'n}kosz and Kamil Ko{\l}ty{\'s}.
\newblock Integral flow decomposition with minimum longest path length.
\newblock {\em European Journal of Operational Research}, 247(2):414--420,
  2015.
\newblock \href {https://doi.org/10.1016/j.ejor.2015.06.012}
  {\path{doi:10.1016/j.ejor.2015.06.012}}.

\bibitem{Shao}
Kingsford Shao, M.
\newblock Theory and {A} heuristic for the minimum path flow decomposition
  problem.
\newblock In {\em IEEE/ACM Transactions on Computational Biology and
  Bioinformatics}, 2017.

\bibitem{SUPPAKITPAISARN2016367}
Vorapong Suppakitpaisarn.
\newblock An approximation algorithm for multiroute flow decomposition.
\newblock {\em Electronic Notes in Discrete Mathematics}, 52:367 -- 374, 2016.
\newblock INOC 2015 -- 7th International Network Optimization Conference.

\bibitem{TomescuGPRKM15}
Alexandru~I. Tomescu, Travis Gagie, Alexandru Popa, Romeo Rizzi, Anna
  Kuosmanen, and Veli M{\"{a}}kinen.
\newblock Explaining a weighted {DAG} with few paths for solving genome-guided
  multi-assembly.
\newblock {\em {IEEE} {ACM} Trans. Comput. Biol. Bioinform.}, 12(6):1345--1354,
  2015.
\newblock \href {https://doi.org/10.1109/TCBB.2015.2418753}
  {\path{doi:10.1109/TCBB.2015.2418753}}.

\bibitem{tomescu2013novel}
Alexandru~I Tomescu, Anna Kuosmanen, Romeo Rizzi, and Veli M{\"a}kinen.
\newblock A novel min-cost flow method for estimating transcript expression
  with rna-seq.
\newblock {\em BMC bioinformatics}, 14(S5):S15, 2013.
\newblock \href {https://doi.org/10.1186/1471-2105-14-S5-S15}
  {\path{doi:10.1186/1471-2105-14-S5-S15}}.

\bibitem{tomescu2017safe}
Alexandru~I. Tomescu and Paul Medvedev.
\newblock Safe and complete contig assembly through omnitigs.
\newblock {\em Journal of Computational Biology}, 24(6):590--602, 2017.
\newblock Preliminary version appeared in RECOMB 2016.

\bibitem{vatinlen2008simple}
Benedicte Vatinlen, Fabrice Chauvet, Philippe Chr{\'e}tienne, and Philippe
  Mahey.
\newblock Simple bounds and greedy algorithms for decomposing a flow into a
  minimal set of paths.
\newblock {\em European Journal of Operational Research}, 185(3):1390--1401,
  2008.
\newblock \href {https://doi.org/10.1016/j.ejor.2006.05.043}
  {\path{doi:10.1016/j.ejor.2006.05.043}}.

\bibitem{citeulike:3614773}
Zhong Wang, Mark Gerstein, and Michael Snyder.
\newblock {RNA-Seq: a revolutionary tool for transcriptomics.}
\newblock {\em Nature Reviews Genetics}, 10(1):57--63, 2009.
\newblock \href {https://doi.org/10.1038/nrg2484} {\path{doi:10.1038/nrg2484}}.

\bibitem{Ref-Sim}
L.~Williams.
\newblock Reference-sim.
\newblock Nov 2021.
\newblock \href {https://doi.org/https://doi.org/10.5281/zenodo.5646910}
  {\path{doi:https://doi.org/10.5281/zenodo.5646910}}.

\bibitem{williams2019rna}
Lucia Williams, Gillian Reynolds, and Brendan Mumey.
\newblock Rna transcript assembly using inexact flows.
\newblock In {\em 2019 IEEE International Conference on Bioinformatics and
  Biomedicine (BIBM)}, pages 1907--1914. IEEE, 2019.

\bibitem{BenchmarkingStudies2}
Lucia Williams, Alexandru~I. Tomescu, and Brendan Mumey.
\newblock {Flow Decomposition with Subpath Constraints}.
\newblock In Alessandra Carbone and Mohammed El-Kebir, editors, {\em 21st
  International Workshop on Algorithms in Bioinformatics (WABI 2021)}, volume
  201 of {\em Leibniz International Proceedings in Informatics (LIPIcs)}, pages
  16:1--16:15, Dagstuhl, Germany, 2021. Schloss Dagstuhl -- Leibniz-Zentrum
  f{\"u}r Informatik.
\newblock URL:
  \url{https://drops.dagstuhl.de/entities/document/10.4230/LIPIcs.WABI.2021.16},
  \href {https://doi.org/10.4230/LIPIcs.WABI.2021.16}
  {\path{doi:10.4230/LIPIcs.WABI.2021.16}}.

\bibitem{findingranges}
Hongyu Zheng, Cong Ma, and Carl Kingsford.
\newblock Deriving ranges of optimal estimated transcript expression due to
  nonidentifiability.
\newblock {\em J. Comput. Biol.}, 29(2):121--139, 2022.
\newblock \href {https://doi.org/10.1089/cmb.2021.0444}
  {\path{doi:10.1089/cmb.2021.0444}}.

\end{thebibliography}


\appendix

\section{Optimal Algorithm for Optimal Concise Representation}\label{second-algo}
\subsubsection*{Data structures}
{We have used the following data structures, as described in the previous work~\cite{optFlowTheory}} 
\begin{itemize}
  \item ${\cal T}_u$ - It represents all the left maximal safe paths ending at u in a Trie structure.
  \item ${\cal L}^+_u$ - It represents the list of safe paths to store the concise representation. The concise representation is in the form 
  {(${\cal P}_1$, ${\cal I}_1$), (${\cal P}_2$, ${\cal I}_2$), ......}, where ${\cal I}_k$ comprises of tuples {($l_k$, $u_k$, $f_k$), ......}
  \item $l_k$ - It represents the left border of an interval.
  \item $u_k$ - It represents the right border of an interval.
  \item $f_k$ - It represents the excess flow value, as stated in Lemma~\ref{lem:excess-flow}, of that interval path from $l_k$ to $u_k$.
\end{itemize}

\subsubsection*{Proposed modifications to the algorithm}
The changes in lines 29-32 highlight that if two different paths ${\cal P}_i$ and ${\cal P}_j$ add the same safe path ${\cal P}_v^*$ to $v^*$, we need to add this to only one of them and the other path must find some other neighbour to extend to, such that the same safe path is not included in the concise representation of more than one paths. Note that if $v^*$ exists ${\cal T}_u$ is added to ${\cal T}_{v^*}$ before processing ${\cal T}_u$, thereafter any deletions from ${\cal T}_u$ will also be carried on ${\cal T}_{v^*}$. Both the cases (whether $v^*$ exists or not) are described together for the sake of simplicity.\\

The other addition to the pseudo code (lines 36-40 and 49-51) involves the removal of unit-length safe paths from the algorithm's output since the unit-length safe paths were also getting added to the solution, and trivially, we know that they will always be safe. Hence this addition reduced the output size and improved the evaluation time of the algorithm.

\normalem
\begin{algorithm}[tbh]
	\caption{Modified Version of Optimal Concise Representation of Safe Paths}
	\label{alg:optConR}

	\DontPrintSemicolon
	\BlankLine

	\begin{multicols}{2}
	Compute Topological Order of $G$\;
    \ForAll{$u\in V$ in topological order}{
        \textsc{Compute-Safe-CompR}$(u)$\;}
    \BlankLine
    \BlankLine
\!\!\!\textsc{Compute-Safe-CompR$(u)$:} \\
    \uIf{$u\notin Sink(G)\cup Source(G)$}
        {$v^*\gets v_{max}(u,\cdot)$}
    \lElse{$v^*\gets null$}

    \If{$u\in Source(G)$}{
        Initialize ${\cal T}_u$ with $u$
    }
    \BlankLine
    \BlankLine
     
       
        \ForAll{$(u,v)\in G, v\neq v^*$}{
            Add $(u,v)$ to ${\cal T}_v$\;
            $x\gets u$ in ${\cal T}_u$, $f_x\gets f(u,v)$\;
            
            \While{$x\neq$ leaf of ${\cal T}_u$ \textbf{and} $f_x-f_{in}(x)+f_{max}(\cdot,x)>0$}{
                Add $e_{max}(\cdot,x)$ to ${\cal T}_v$\;
                $f_x\gets f_x-f_{in}(x)+f_{max}(\cdot,x)$\;
                $x\gets v_{max}(\cdot,x)$ in ${\cal T}_u$\; 
            }
           
            Add $(\emptyset,\{(x,v,f_x)\})$ to ${\cal L}^+_v$\;
            Push $v$ to $Mark[x]$\;
            Add $x$ to $\cal M$\;
        }
 
     \If{$v^*\neq null$}{
        Add ${\cal T}_u$ as child of $v^*$ in ${\cal T}_{v^*}$\;        
        Future changes to ${\cal T}_u$ will change ${\cal T}_{v^*}$     
            
            }

    \ForAll{$(p_k,I_k)\in {\cal L}^+_u$}{
        $(l_i,u,f_i)\gets$ Last of $I_k$, 
        $x\gets l_i$ in ${\cal T}_u$\;
        \lIf{$v^*= null$}{$f_x\gets -\infty$}
        \lElse {$f_x\gets f_i-f_{out}(u)+f(u,v^*)$}
        \BlankLine        

         \While{$f_x\leq 0$ \textbf{and} $Mark[x]=\emptyset$ \textbf{and} $x\neq u$}{
                    $y\gets $ Parent of $x$ in ${\cal T}_{u}$\;
                    \color{blue} 
                    \uIf{$x$ is a leaf in ${\cal T}_{u}$}{
                    $f_x\gets f_x+f_{in}(y)-f(x,y)$\;
                    Remove $(x,y)$ from ${\cal T}_{u}$\;
                    }
                    \lElse{
                        break
                    }
                    \color{black}
                    $x\gets y$\;
                }
                
                $p_x\gets$ Path from $l_i$ to $x$ in ${\cal T}_{u}$\; 
                $p_k\gets p_k\cup \{p_x\setminus \{x\}\}$\;  
                
                \color{blue}
                \If{$p_k\neq\emptyset$ \textbf{and} $x$ is root of ${\cal T}_{u}$ \textbf{and} \\ $x$ was parent of $l_i$ in ${\cal T}_{u}$}{
                    Pop from $I_k$\;
                    \lIf{$I_k\neq\emptyset$}{Add $(p_k ,I_k)$ to \textsc{Sol}}
                    Clear $p_k,I_k$\;
                }
                \uIf{$f_x>0$ \textbf{and} $x$ is a leaf in ${\cal T}_{u}$}{
                \color{black}
                    \lIf{$l_i\neq x$}{
                        Add $(x,v^*,f_x)$ to $I_k$
                    }\lElse{
                        Last of $I_k\gets (l_i,v^*,f_x)$
                    }    
                    Add $(p_k,I_k)$ to ${\cal L}^+_{v^*}$\;
                }
                \color{black}
                \uElseIf{Mark$[x]\neq \emptyset$}{
                    $v\gets $ Pop from Mark$[x]$\;
                    $(\emptyset,I_v)\gets $ Pop from ${\cal L}^+_v$\;
                    Add $(p_k,I_v\cup I_k)$ to ${\cal L}^+_v$\;
                }
                \color{blue}
                \ElseIf{$I_k\neq\emptyset$}{
                    $p_u\gets$ Path from $x$ to $u$ in ${\cal T}_{u}$\; 
                    Add $(p_k\cup p_u ,I_k)$ to \textsc{Sol}\;
                }
                \color{black}
    }            
    \lForAll{$x\in {\cal M}$}{Clear Mark$[x]$}       
    Clear ${\cal M}$\;
    
   \end{multicols}
\end{algorithm}

\ULforem

\section{Algorithm for Optimal Representation}\label{Third-algo-code}

\subsection{Data Structures}
We have employed the identical data structures as those utilized previously in~\cite{optFlowTheory}.
\begin{itemize}
\item \textbf{Unique maximum incoming and outgoing forests $F_i$ and $F_o$} \\ They represent the unique maximum incoming/outgoing forest.
\item \textbf{Level Ancestors $LA_i$ and $LA_o$} \\ The level ancestors are used to calculate the ancestor of a node in $F_i$/$F_o$ at some depth d.
\item \textbf{Cumulative Losses $c_i$ and $c_o$} \\ The cumulative loss for a node in $c_i/c_o$ is equal to the loss of excess flow when the path is left/right extended from the node to the root in $F_i/F_o$.
\end{itemize}

\subsection{Algorithm}

All the data structures needed are precomputed before actually computing the safe paths. The maximum incoming/outgoing edge from each node is added to $F_i/F_o$, if it exists. The nodes are traversed in topological order, starting from the root to the leaves, for calculating $c_i$ and $c_o$. The values for subsequent nodes are calculated by subtracting the excess flow loss, as given in Lemma~\ref{lem:excess-flow}, from the value of the parent node. We use~\cite{LevelAncestor} for calculating the data structures $LA_i$ and $LA_o$.

For enumerating the non-trivial safe paths, select every edge which is not a unique maximum incoming edge, say $e(u,v)$. We then apply the two-pointer approach to find all the maximal safe paths with the selected edge as the representative edge. To get the first path, a binary search is performed in the ancestors of $u$ in $F_i$. This gives us the left end, say $l$ of the path, which is selected such that $f(u,v)$ is greater than the loss of excess flow in extending the path from $u$ to $l$. The binary search is performed on the depth of the node in $F$, and the node at that depth is calculated using $LA$. For this, the loss in flow is $c_i[l]-c_i[u]$. On a similar note, the right end, say $r$ is searched in $F_o$ using the excess flow value of $f(u,v)-c_i[l]+c_i[u]$. Then we add the path $<l,...u,v....,r>$ to the solution along with the excess flow value of $f(u,v)-c_i[l]+c_i[u]-c_o[r]+c_o[v]$. After this is done, we have got the first path. For finding the next path, the right end is extended by an edge. Then this right extended path is used to search for the left end. As before, the right end can be calculated in a similar fashion. The right extension may not be possible due to either the right end reaching the root or the right extended path becoming unsafe. In such a case, we move to the next edge. Only paths $p$ with $|p| > 1$ are added to the solution. Note that we also check whether a path is not left extendable before adding it to the solution because it may be a subpath of another safe path with the representative edge $(l',l)$. This case is possible if $(l',l)$ is a maximum incoming edge of $l$ but not the unique maximum incoming edge of $l$.

For enumerating the trivial safe paths, start from each leaf in $F_i$, say $u$ and parent of $u$, say $v$. With $u$ as the right end and $e(u,v)$ as the representative edge, we find the left end $x$ for the first safe path. Only in the case the right extension of the path $<x,...v,u>$ by an edge is not possible or makes the path unsafe it is added to the solution along with the flow value of $f(v,u)-c_i[x]+c_i[v]$. To find the next path, extend the left end by one edge. This new left end, say $y$, is used to search for the right end. Since we need to search in the descendants of $y$, and there is no direct way to search in the descendants, instead we search on the ancestors of the previous right end. Using this approach, the right end can be used to find the left end again. To overcome the challenge of duplicates in the solution, we mark the right end of every path we get. This ensures that even if the same node is visited through another leaf, we do not add the path again in the solution. We move on to the next leaf and this process is repeated when the left extension of the path is not further possible.

The following notations have been used in the pseudocode~\ref{alg:optR} provided -
\begin{itemize}
  \item $LeftExtend(u,excess)$ -  It returns the node obtained by maximally extending the path along the ancestors of $u$ in $F_i$ such that the loss of flow by the extension is less than $excess$.
  \item $RightExtendO(u,excess)$ - It returns the node obtained by extending the path along the ancestors of $u$ in $F_o$ such that the loss of flow by the extension is less than $excess$.
  \item $RightExtendI(u)$ - It returns the deepest descendant $v$ of $u$ in $F_i$ such that the path $<u,...,v>$ is safe.
  \item $IsLeftExtendable(u,excess)$ - It returns $true$ if the path can be extended on the left of $u$ with a flow loss less than $excess$, else $false$.
  \item $IsRightExtendable(u,excess)$ - It returns $true$ if the path can be extended on the right of $u$ with a flow loss less than $excess$, else $false$.
\end{itemize}

\normalem

\begin{algorithm}[tbh]
	\caption{Optimal Representation of Safe Paths}
	\label{alg:optR}
        \DontPrintSemicolon
\begin{multicols}{2}
\tcp{Compute $F_i,F_o,c_i[\cdot], c_o[\cdot],LA_i,LA_o$}


\ForAll {$u \in$ leaves of $F_i$}{
    \While {$marked[u]$ is $false$} {
        $marked[u] \gets true$ \\
        $v \gets$ Parent of $u$ in $F_i$ \\
        $x \gets $$LeftExtend(v,f(u,v))$\\
            $flow \gets f(v,u)-c_i[x]+c_i[v]$ \\ 
        \If {$not$ $IsLeftExtendable(x,flow)$ and $not$ $IsRightExtendable(u,flow)$ and $x\neq v$} {
            Add path $(x,v,u,u,flow)$ 
        }
        \lIf {$x$ is $root$ in $F_i$} {\textbf{break}}
        $y \gets $ Parent of $x$ \\
        $u \gets $$RightExtendI(y)$ 
    }
}

\ForAll {$e(u,v)\in E\setminus F_i$}{
    $excess \gets f(u,v)$ \\
    \While {$excess>0$} {
        $l \gets LeftExtend(u,excess)$\\
        $excess \gets f(u,v) -c_i[l]+c_i[u]$ \\
        $r \gets RightExtendO(v,excess)$\\
            $flow \gets excess - c_o[r] + c_o[v]$ \\
        \If {($l \neq u$ or $r \neq v$) and 
        $not$ $IsLeftExtendable(l,flow)$}{
            Add path $(l,u,v,r,flow)$ 
        }
        \lIf {$r$ is $root$} {\textbf{break}}
        $next \gets$ Parent of $r$ in $F_o$ \\
        $excess \gets f(u,v) - c_o[next]+c_o[v]$
    }
}

\vspace*{0.01cm}
\end{multicols}

\end{algorithm}

\ULforem

\section{Actual Performance Evaluation}
\label{apn:extraTables}
Consider \Cref{tab:actP}, \texttt{RawRep$_o$} and \texttt{ConRep$_o$} represent the algorithms used by~\cite{KhanMMLA22} without removing duplicates, suffixes, and prefixes using AC Trie. Clearly, the memory requirements decrease as compared to \texttt{RawRep} and \texttt{ConRep} because AC Trie is not being built. However, the time requirement can both increase or decrease. This can be explained using the relative sizes of the generated output (see \Cref{tab:actS}) because only in the case of Ref-Sim and Simulation the size decreases drastically ($>$3x) where the difference in time required to print the output may dominate the time required by AC Trie.

\begin{table}[!h]
\centering
\begin{tabular}{|c|c|c|c|c|c|c|c|} \hline
Algorithm & Parameter & Zebrafish & Mouse & Human & Salmon & Ref-Sim & Simulation \\ \hline
\multirow{2}{*}{\texttt{RawRep$_o$}} & Time & \textbf{7.29} & \textbf{9.02} & \textbf{10.41} & 200.63 & 0.72 & 13.23 \\
 & Memory & \textbf{1.73} & \textbf{1.77} & \textbf{1.76} & \textbf{1.76} & \textbf{1.82} & 3.86 \\ \hline
\multirow{2}{*}{\texttt{RawRep}} & Time & 8.52 & 10.16 & 11.5 & 210.72 & 0.64 & 9.6 \\
 & Memory & 1.89 & 2.09 & 1.92 & 1.97 & 2.34 & 3.89 \\ \hline
\multirow{2}{*}{\texttt{ConRep$_o$}} & Time & 7.42 & 9.32 & 10.93 & 205.91 & 0.76 & 15.22 \\
 & Memory & 1.74 & 1.78 & 1.77 & 1.77 & 1.83 & 3.92 \\ \hline
\multirow{2}{*}{\texttt{ConRep}} & Time & 9.03 & 11.3 & 12.91 & 249.19 & 0.98 & 10.54 \\
 & Memory & 1.85 & 2.11 & 1.89 & 2.0 & 3.13 & 3.97 \\ \hline
\multirow{2}{*}{\texttt{OptRawRep}} & Time & 8.19 & 9.31 & 10.72 & \textbf{182.27} & 0.5 & \textbf{6.05} \\
 & Memory & 1.8 & 1.86 & 1.86 & 1.86 & 1.85 & \textbf{2.1} \\ \hline
\multirow{2}{*}{\texttt{OptConRep}} & Time & 8.48 & 10.04 & 11.66 & 197.9 & 0.57 & 7.37 \\
 & Memory & 1.82 & 1.87 & 1.87 & 1.84 & 1.86 & 2.25 \\ \hline
\multirow{2}{*}{\texttt{OptRep}} & Time & 8.52 & 9.71 & 11.21 & 183.0 & \textbf{0.48} & 7.35 \\
 & Memory & 1.88 & 1.94 & 1.91 & 1.91 & 2.09 & 2.35 \\ \hline

   \end{tabular} 
    \caption{Time ($sec$) and Memory ($MB$) of algorithms without removing duplicates.}\label{tab:actP}
\end{table}

\begin{table}[!h]
\centering
\begin{tabular}{|c|c|c|c|c|c|c|} \hline
Algorithm & Zebrafish & Mouse & Human & Salmon & Ref-Sim & Simulation \\ \hline
\texttt{RawRep$_o$} & 70.8 & 103.39 & 111.48 & 2858.33 & 14.02 & 177.41 \\ \hline
\texttt{RawRep} & 49.83 & 60.32 & 66.71 & 1351.57 & 3.89 & \textbf{20.5} \\ \hline
\texttt{ConRep$_o$} & 63.81 & 92.08 & 103.58 & 2396.35 & 12.18 & 240.47 \\ \hline
\texttt{ConRep} & 57.29 & 85.07 & 95.9 & 2225.37 & 11.26 & 31.4 \\ \hline
\texttt{OptRawRep} & 49.83 & 60.32 & 66.71 & 1351.57 & 3.89 & \textbf{20.5} \\ \hline
\texttt{OptConRep} & 46.94 & 58.69 & 66.87 & 1269.3 & 4.14 & 26.72 \\ \hline
\texttt{OptRep} & \textbf{21.22} & \textbf{28.98} & \textbf{35.14} & \textbf{719.17} & \textbf{2.29} & 22.05 \\ \hline
   \end{tabular} 
    \caption{Output Size ($MB$) of algorithms without removing duplicates.}\label{tab:actS}
\end{table}

\begin{table}[!h]
\centering
\begin{tabular}{|c|c|c|c|c|c|c|c|} \hline
Algorithm & Parameter & Zebrafish & Mouse & Human & Salmon & Ref-Sim & Simulation \\ \hline
\multirow{3}{*}{\texttt{OptRawRep}} & Time & 8.19 & 9.31 & 10.72 & 182.27 & 0.5 & \textbf{6.05} \\
 & Memory & 1.8 & 1.86 & 1.86 & 1.86 & 1.85 & \textbf{2.1} \\
 & Output Size & 49.83 & 60.32 & 66.71 & 1351.57 & 3.89 & 20.5 \\ \hline
\multirow{3}{*}{\texttt{OptConRep}} & Time & 8.48 & 10.04 & 11.66 & 197.9 & 0.57 & 7.37 \\
 & Memory & 1.82 & 1.87 & 1.87 & 1.84 & 1.86 & 2.25 \\
 & Output Size & 46.94 & 58.69 & 66.87 & 1269.3 & 4.14 & 26.72 \\ \hline
\multirow{3}{*}{\texttt{OptConRep$^\#$}} & Time & \textbf{7.83} & \textbf{9.18} & \textbf{10.46} & \textbf{178.74} & 0.51 & 6.81 \\
 & Memory & \textbf{1.77} & \textbf{1.81} & \textbf{1.81} & \textbf{1.8} & \textbf{1.8} & 2.19 \\
 & Output Size & 41.68 & 51.3 & 57.94 & 1093.53 & 3.48 & 19.88 \\ \hline
\multirow{3}{*}{\texttt{OptRep}} & Time & 8.52 & 9.71 & 11.21 & 183.02 & 0.48 & 7.35 \\
 & Memory & 1.88 & 1.94 & 1.91 & 1.91 & 2.09 & 2.35 \\
 & Output Size & 21.22 & 28.98 & 35.14 & 719.17 & 2.29 & 22.05 \\ \hline
\multirow{3}{*}{\texttt{OptRep$^\#$}} & Time & 8.45 & 9.56 & 11.08 & 179.58 & \textbf{0.47} & 7.27 \\
 & Memory & 1.88 & 1.93 & 1.9 & 1.9 & 2.09 & 2.36 \\
 & Output Size & \textbf{17.79} & \textbf{24.36} & \textbf{29.47} & \textbf{600.3} & \textbf{1.95} & \textbf{17.56} \\ \hline
   \end{tabular} 
    \caption{Time ($sec$), Memory ($MB$) and Output Size ($MB$) of algorithms using heuristics.}\label{tab:result-size-1}
\end{table}

\section{Omitted Plots}
\label{apn:extraPlots}

\begin{figure}[H]

\begin{subfigure}[t]{.34\textwidth}
\includegraphics[trim=5 7 10 7, clip,scale=.14]{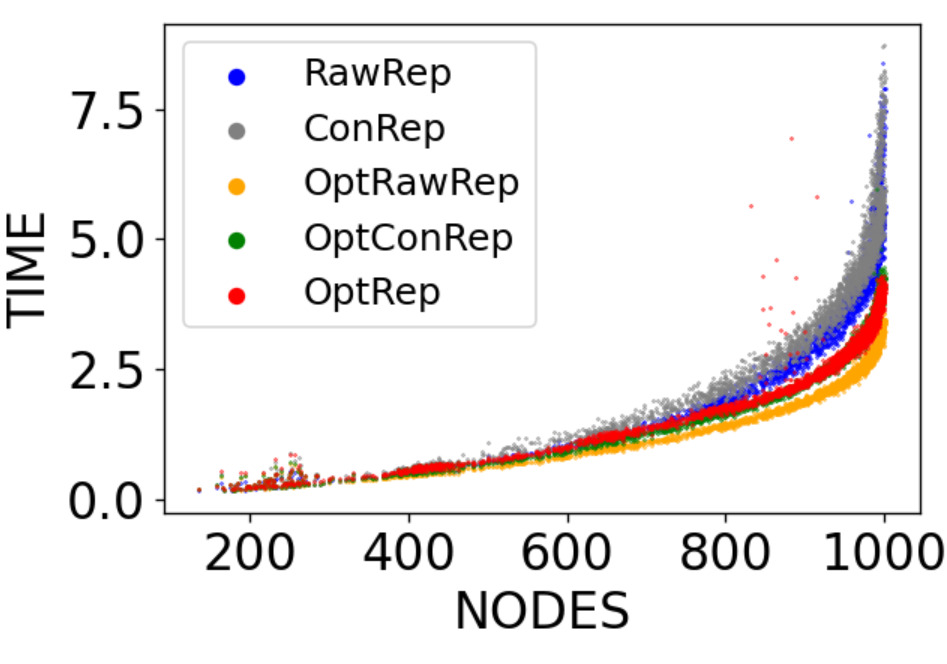}
    \caption{Time ($sec$)}
    \label{fig:timeVnode}
\end{subfigure}
\hfill
\begin{subfigure}[t]{.32\textwidth}
    \includegraphics[trim=5 7 10 7, clip,scale=.14]{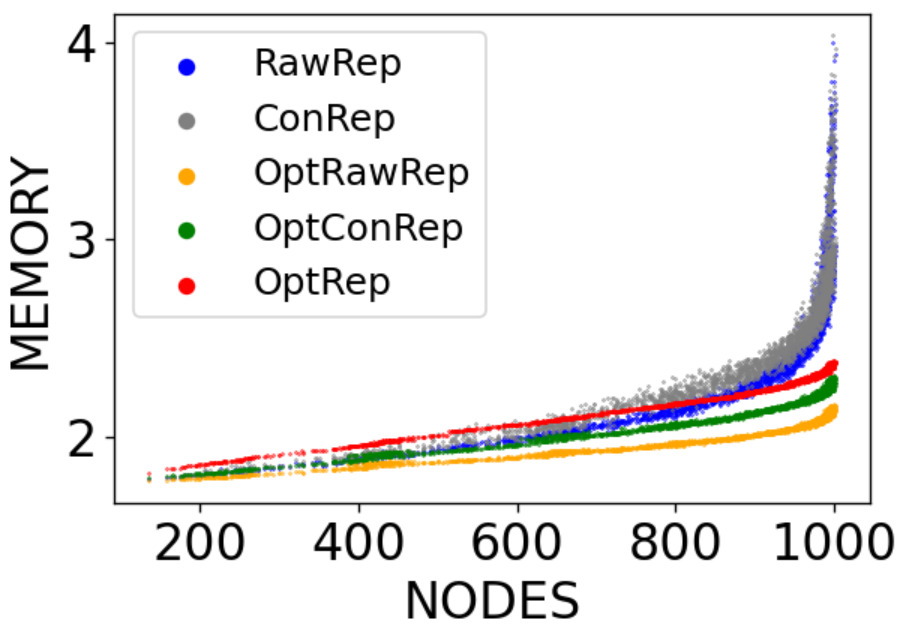}
    \caption{Memory ($MB$)}
    \label{fig:memVnode}
\end{subfigure}
\hfill
\begin{subfigure}[t]{.32\textwidth}
    \includegraphics[trim=5 7 10 7, clip,scale=.14]{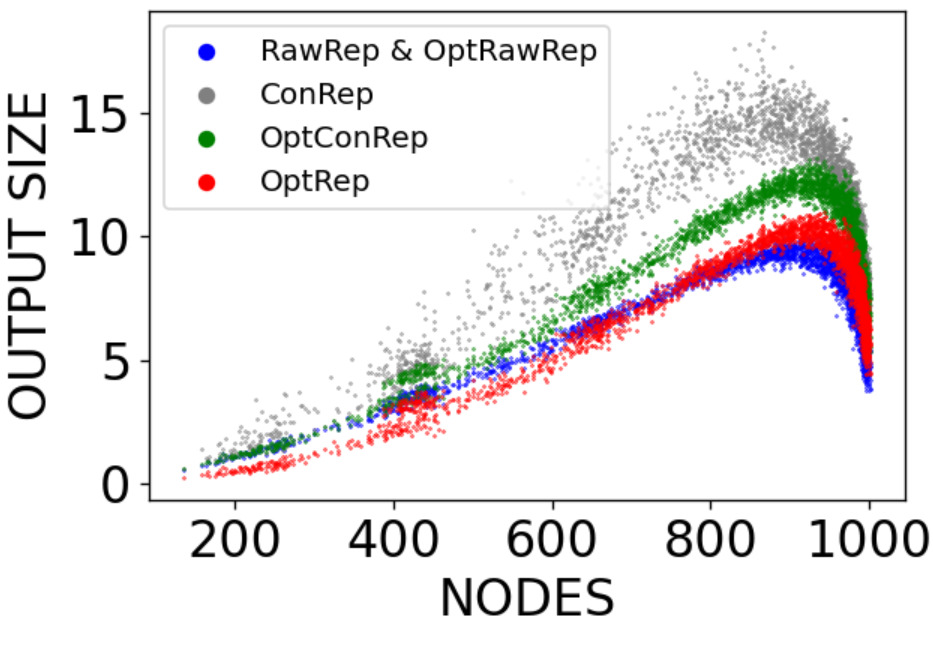}
    \caption{Output Size ($MB$)}
    \label{fig:sizeVnode}
\end{subfigure}
\caption{Performance measures of the algorithms for Simulation w.r.t. number of nodes.}
\label{fig:plotVnodes}
\end{figure}

\begin{figure}[h]
    \centering
    \includegraphics[trim=5 7 10 7, clip,scale=.2]{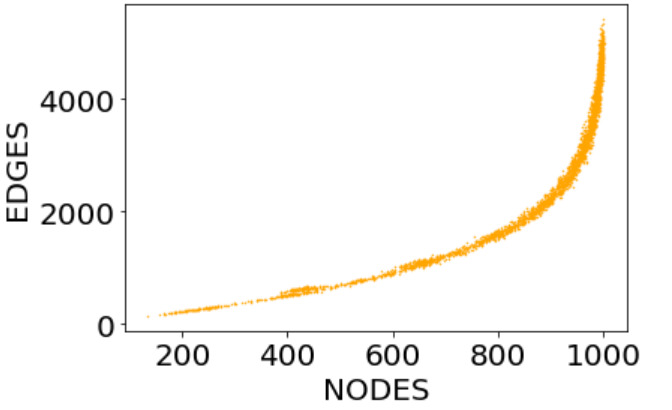}
    \caption{Scatter plot for edges vs nodes in the Simulation dataset.}
    \label{fig:edge-vs-nodes}
\end{figure}

\begin{figure}[H]
\begin{subfigure}[t]{.32\textwidth}
\includegraphics[trim=5 7 10 7, clip,scale=.14]{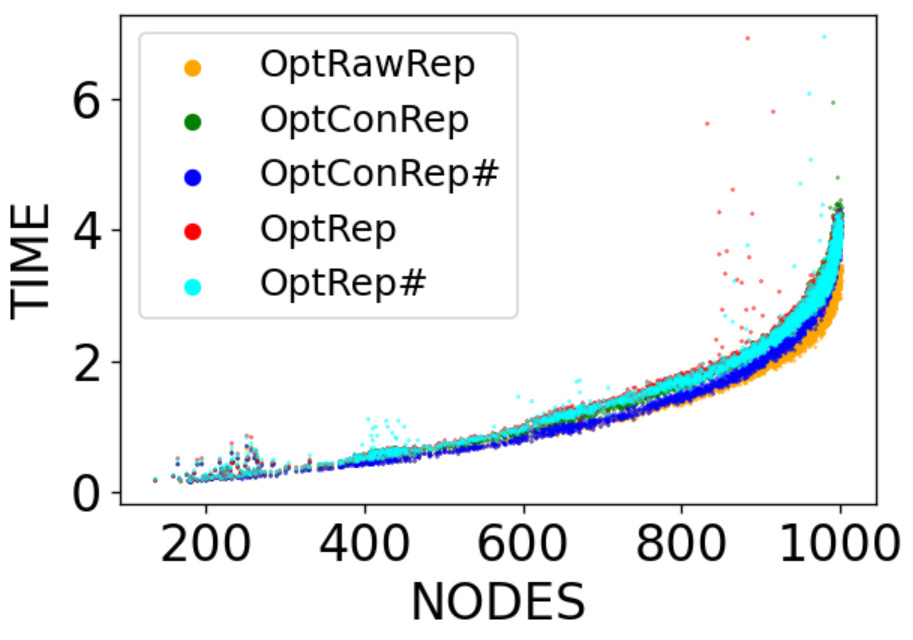}
    \caption{Time ($sec$)}
    \label{fig:timeVnode-heuristics}
\end{subfigure}
\hfill
\begin{subfigure}[t]{.33\textwidth}
    \includegraphics[trim=5 7 10 7, clip,scale=.14]{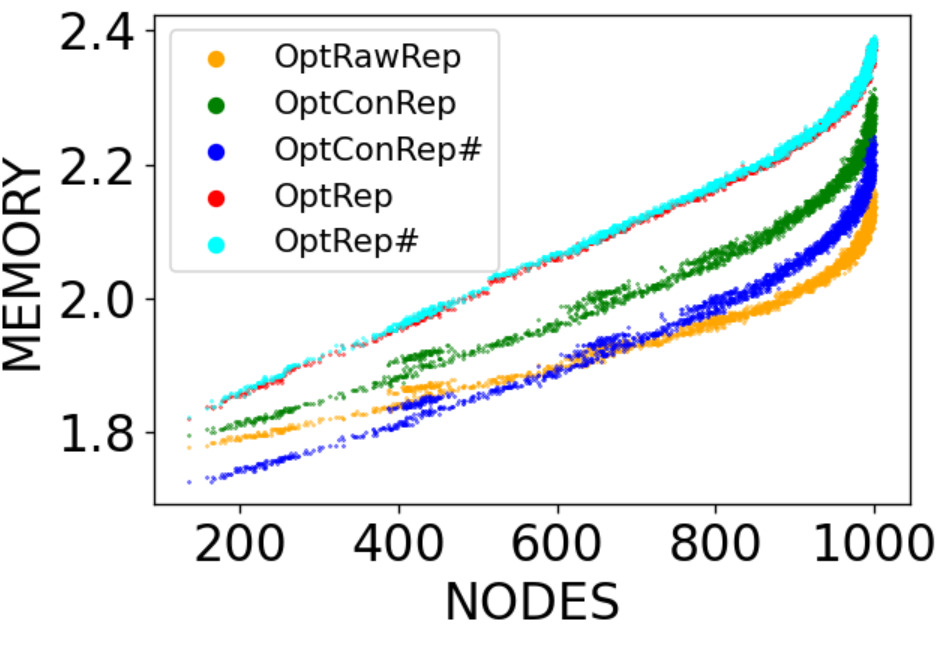}
    \caption{Memory ($MB$)}
    \label{fig:memVnode-heuristics}
\end{subfigure}
\hfill
\begin{subfigure}[t]{.32\textwidth}
    \includegraphics[trim=5 7 10 7, clip,scale=.14]{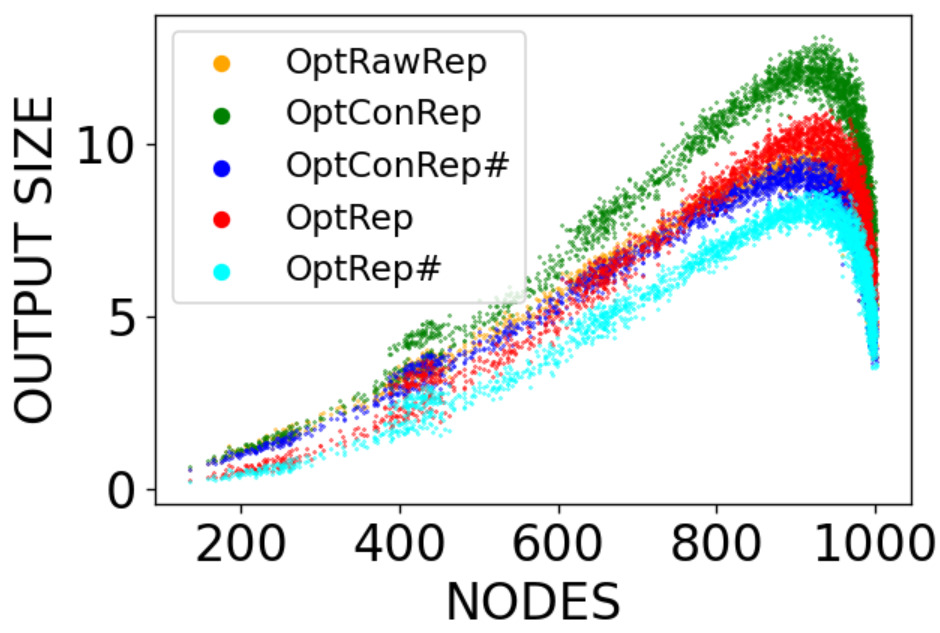}
    \caption{Output Size ($MB$)}
    \label{fig:sizeVnode-heuristics}
\end{subfigure}
\caption{Performance measures of the new algorithms for Simulation w.r.t. number of nodes.}
\label{fig:allVnodesNew}
\end{figure}

\begin{figure}[H]
\begin{subfigure}[t]{0.32\textwidth}
\includegraphics[trim=5 7 10 7, clip,scale=.19]{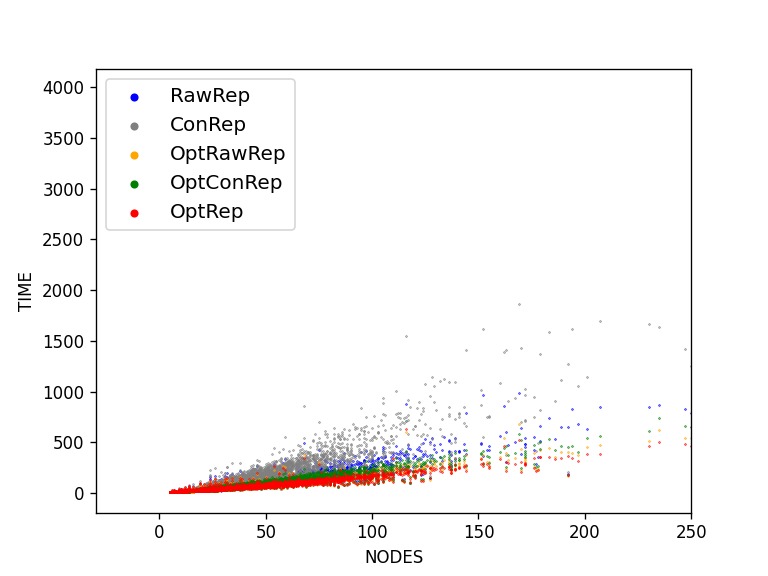}
    \caption{Time ($ms$) vs Nodes}
    \label{fig:timeVnode-refsim}
\end{subfigure}
\hfill
\begin{subfigure}[t]{0.33\textwidth}
    \includegraphics[trim=5 7 10 7, clip,scale=.19]{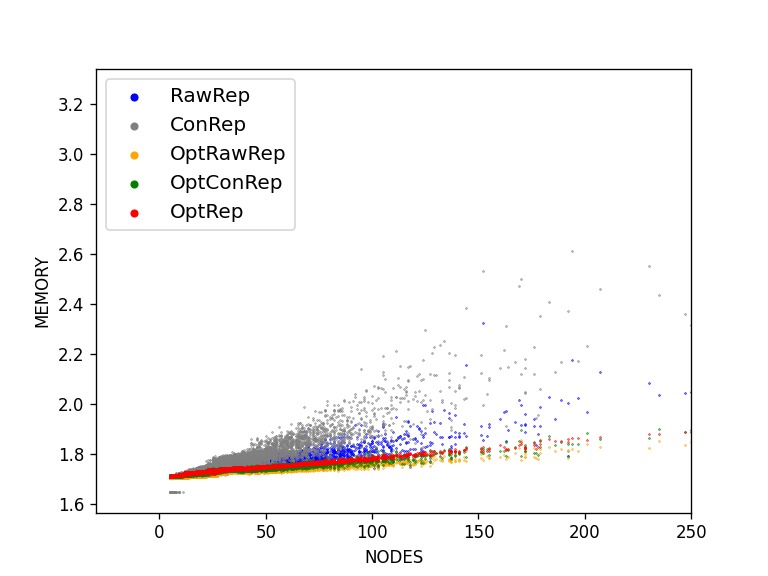}
    \caption{Memory ($MB$) vs Nodes}
    \label{fig:memVnode-refsim}
\end{subfigure}
\hfill
\begin{subfigure}[t]{0.32\textwidth}
    \includegraphics[trim=5 7 10 7, clip,scale=.19]{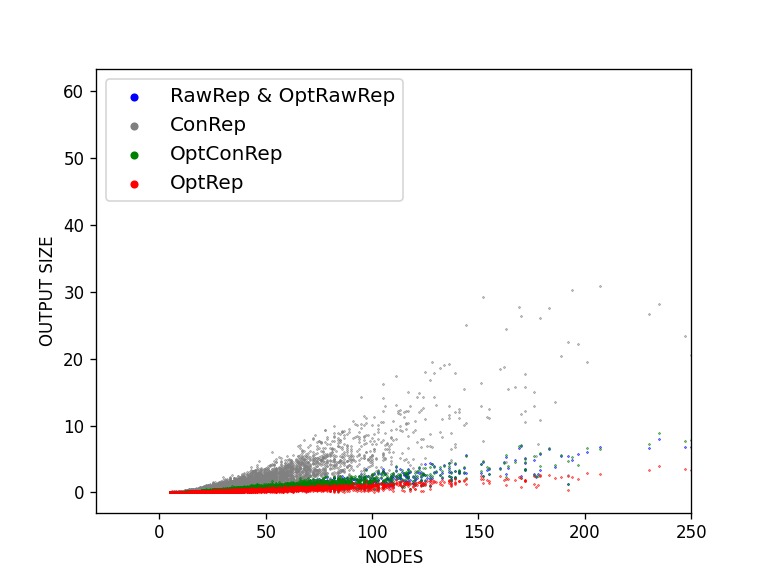}
    \caption{Output Size ($MB$) vs Nodes}
    \label{fig:sizeVnode-refsim}
\end{subfigure}

\begin{subfigure}[t]{0.32\textwidth}
\includegraphics[trim=5 7 10 7, clip,scale=.19]{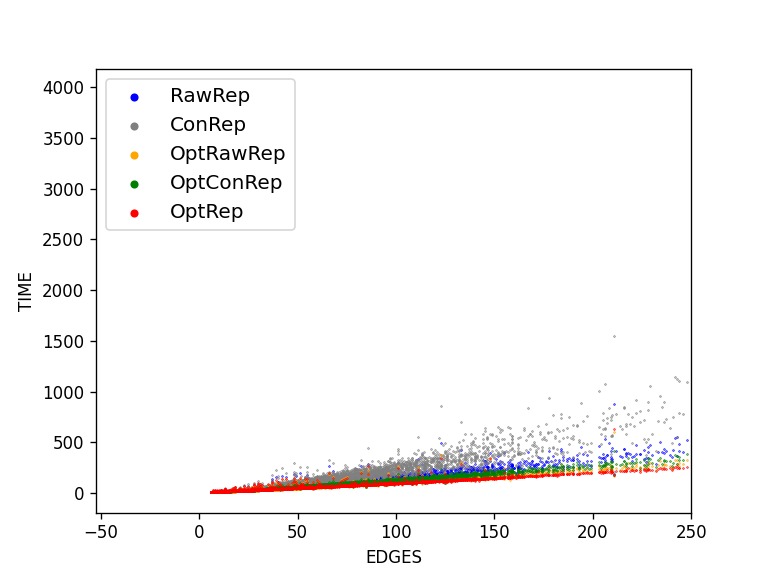}
    \caption{Time ($ms$) vs Edges}
    \label{fig:timeVedge-refsim}
\end{subfigure}
\hfill
\begin{subfigure}[t]{0.33\textwidth}
    \includegraphics[trim=5 7 10 7, clip,scale=.19]{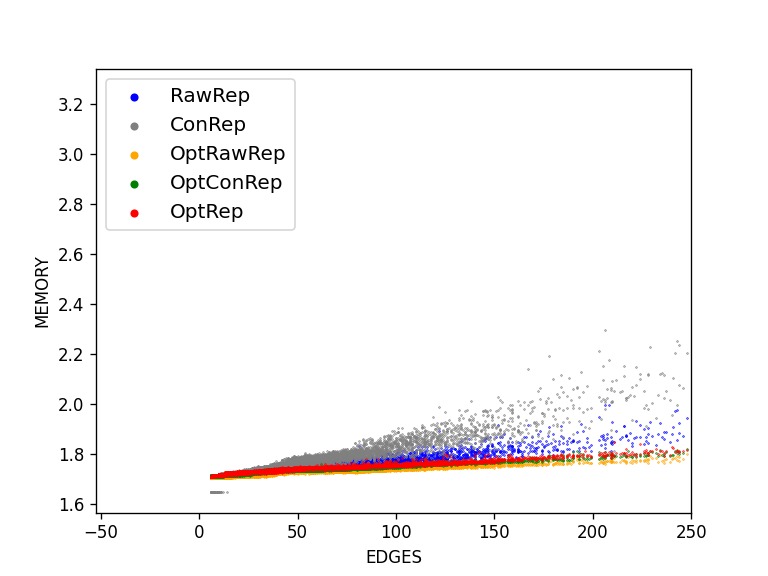}
    \caption{Memory ($MB$) vs Edges}
    \label{fig:memVedge-refsim}
\end{subfigure}
\hfill
\begin{subfigure}[t]{0.32\textwidth}
    \includegraphics[trim=5 7 10 7, clip,scale=.19]{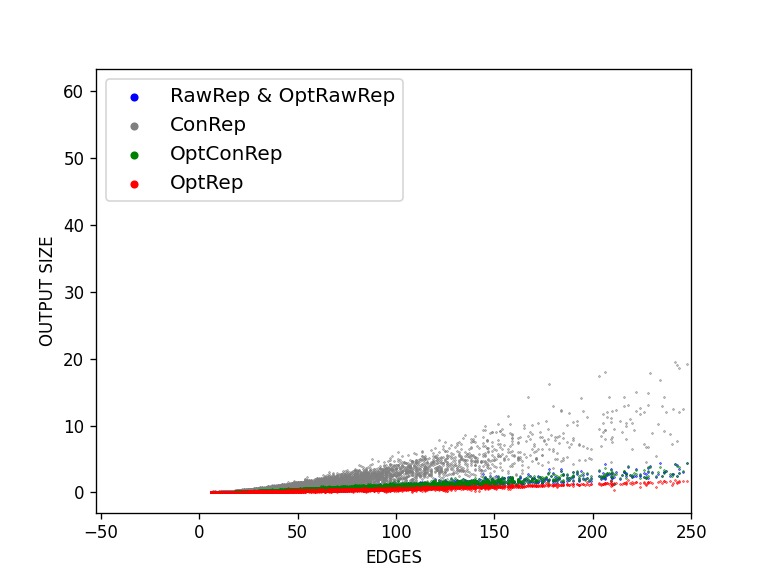}
    \caption{Output Size ($MB$) vs Edges}
    \label{fig:sizeVedge-refsim}
\end{subfigure}
\caption{Performance measures of the algorithms for Ref-Sim w.r.t. nodes and edges.}
\label{fig:RefSim}
\end{figure}

\begin{figure}[H]
\begin{subfigure}[t]{0.45\textwidth}
\includegraphics[trim=5 7 10 7, clip,scale=.25]{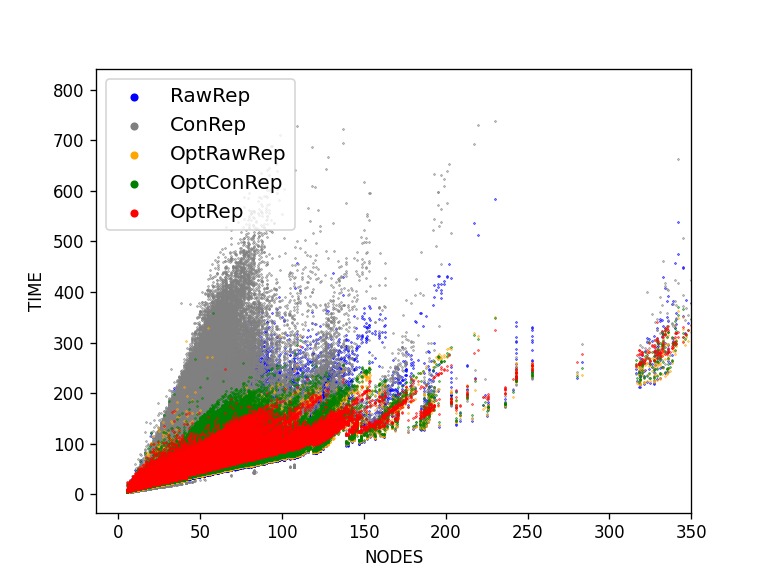}
    \caption{Time ($ms$) vs Nodes}
    \label{fig:timeVnode-catfish}
\end{subfigure}
\hfill
\begin{subfigure}[t]{0.45\textwidth}
    \includegraphics[trim=5 7 10 7, clip,scale=.25]{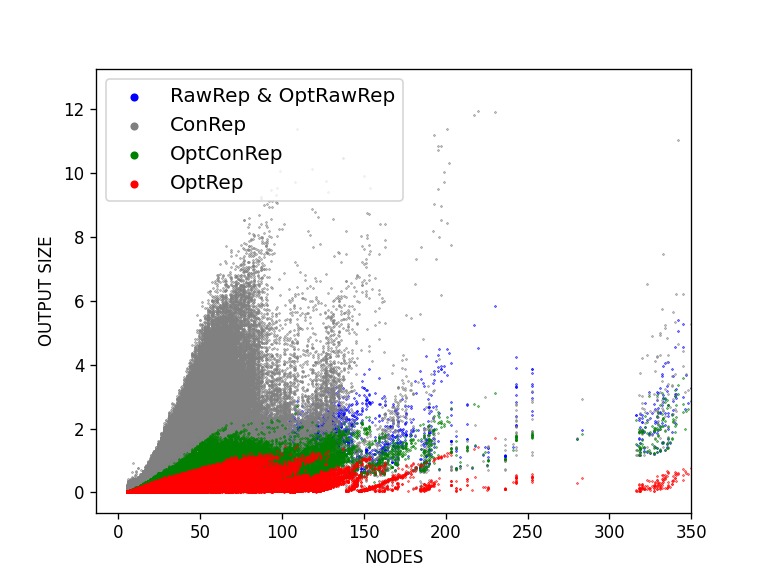}
    \caption{Output Size ($MB$) vs Nodes}
    \label{fig:sizeVnode-catfish}
\end{subfigure}

\begin{subfigure}[t]{0.45\textwidth}
\includegraphics[trim=5 7 10 7, clip,scale=.25]{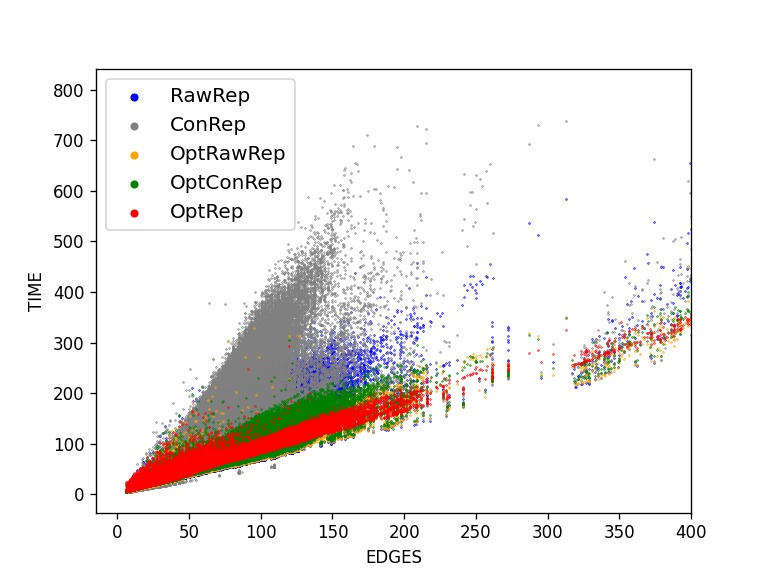}
    \caption{Time ($ms$) vs Edges}
    \label{fig:timeVedge-catfish}
\end{subfigure}
\hfill
\begin{subfigure}[t]{0.45\textwidth}
    \includegraphics[trim=5 7 10 7, clip,scale=.25]{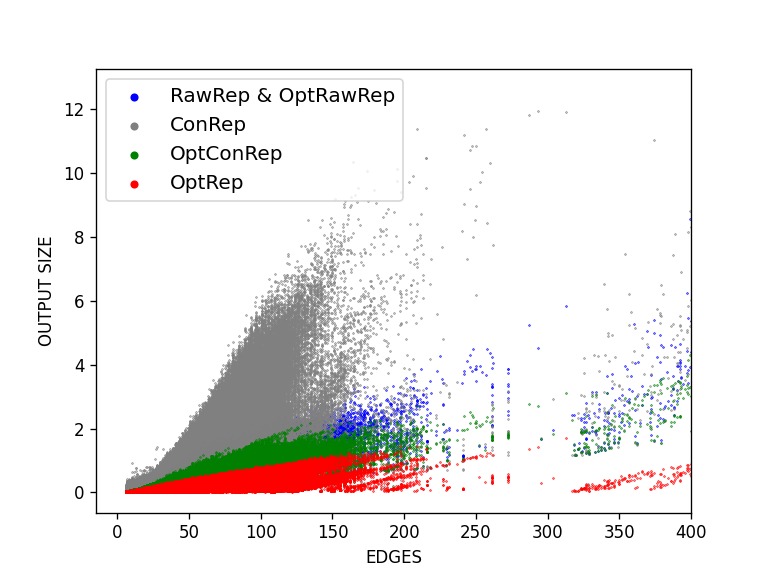}
    \caption{Output Size ($MB$) vs Edges}
    \label{fig:sizeVedge-catfish}
\end{subfigure}
\caption{Performance measures of the algorithms for Catfish w.r.t. nodes and edges.}
\label{fig:catFish}
\end{figure}

\begin{figure}[H]
\begin{subfigure}[t]{0.45\textwidth}
\includegraphics[trim=5 7 10 7, clip,scale=.25]{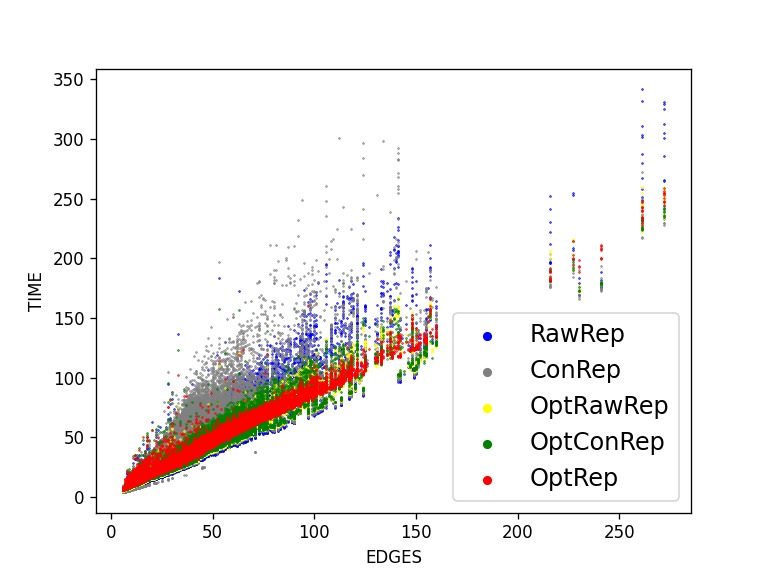}
    \caption{Time ($ms$)}
    \label{fig:timeVnode-zebrafish}
\end{subfigure}
\hfill
\begin{subfigure}[t]{0.45\textwidth}
\includegraphics[trim=5 7 10 7, clip,scale=.25]{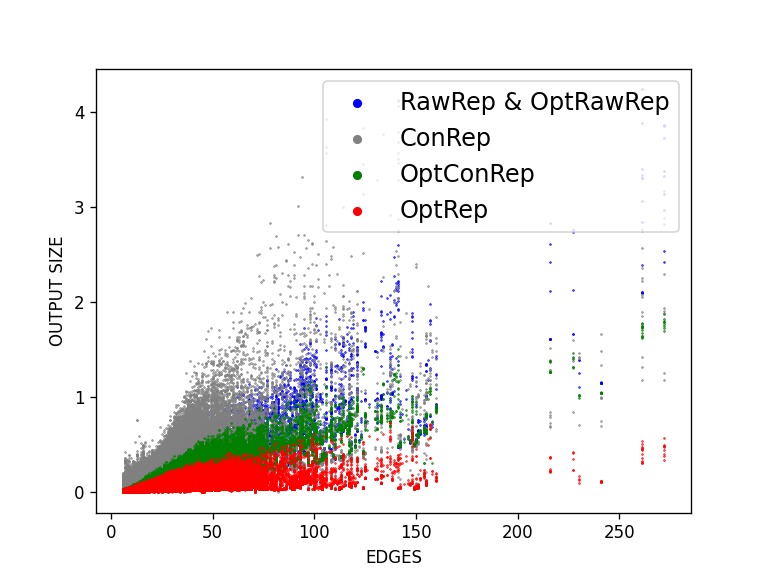}
    \caption{Output Size ($MB$)}
    \label{fig:sizeVnode-zebrafish}
\end{subfigure}
\hfill
\caption{Performance measures of the algorithms for Catfish (Zebrafish) w.r.t. number of edges.}
\label{fig:catFish1}
\end{figure}

\begin{figure}[H]
\begin{subfigure}[t]{0.45\textwidth}
    \includegraphics[trim=5 7 10 7, clip,scale=.25]{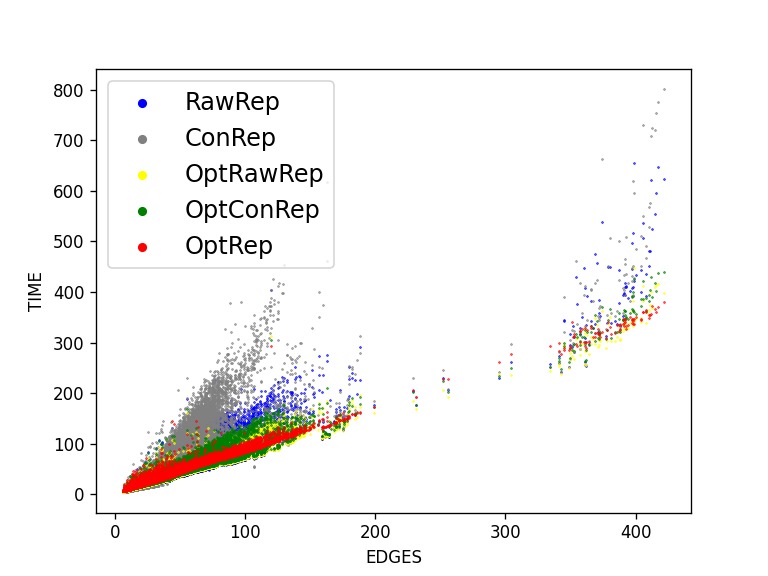}
    \caption{Time ($ms$)}
    \label{fig:timeVnode-mouse}
\end{subfigure}
\hfill
\begin{subfigure}[t]{0.45\textwidth}
    \includegraphics[trim=5 7 10 7, clip,scale=.25]{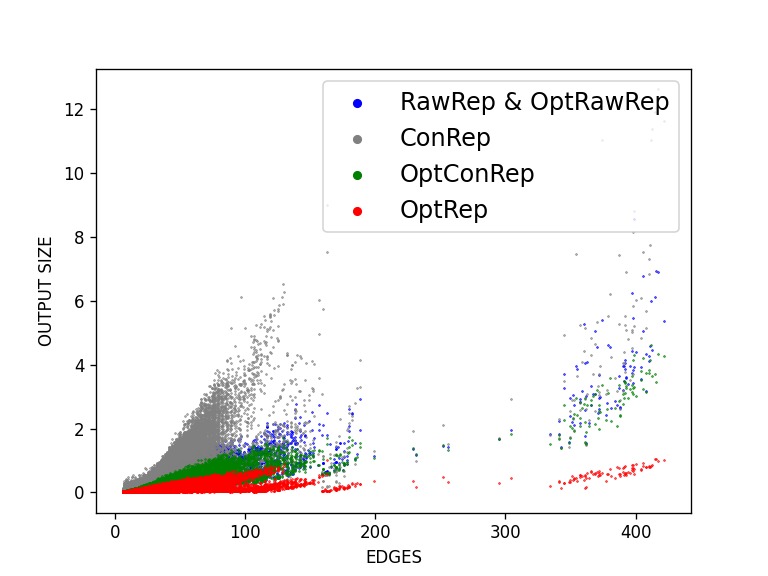}
    \caption{Output Size ($MB$)}
    \label{fig:sizeVnode-mouse}
\end{subfigure}
\hfill
\caption{Performance measures of the algorithms for Catfish (Mouse) w.r.t. number of edges.}
\label{fig:catFish2}
\end{figure}

\begin{figure}[H]
\begin{subfigure}[t]{0.45\textwidth}
    \includegraphics[trim=5 7 10 7, clip,scale=.25]{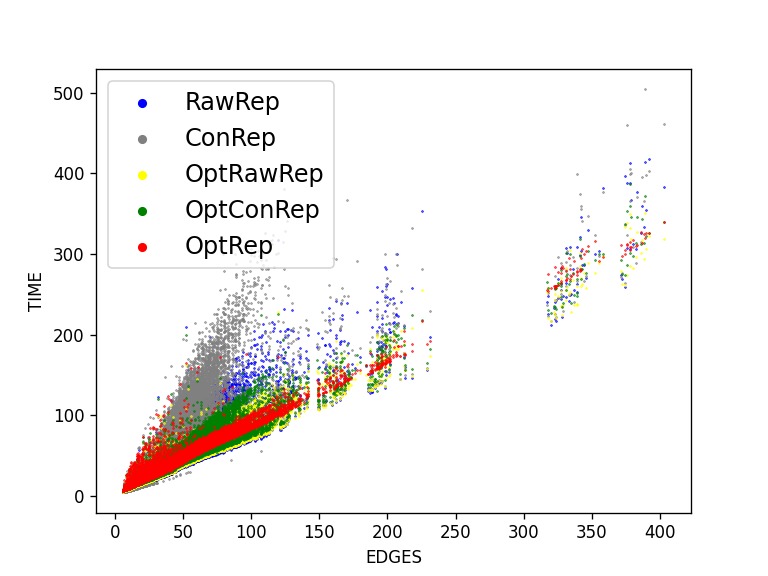}
    \caption{Time ($ms$)}
    \label{fig:timeVnode-human}
\end{subfigure}
\hfill
\begin{subfigure}[t]{0.45\textwidth}
    \includegraphics[trim=5 7 10 7, clip,scale=.25]{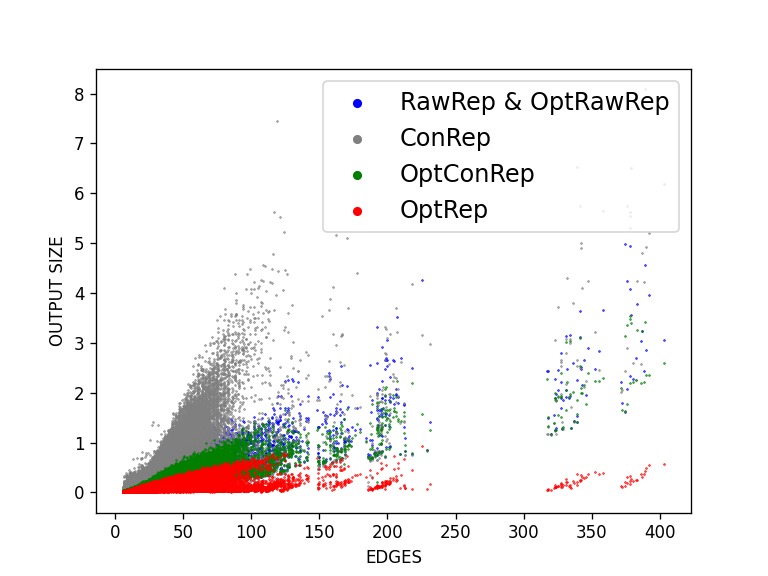}
    \caption{Output Size ($MB$)}
    \label{fig:sizeVnode-human}
\end{subfigure}
\hfill
\caption{Performance measures of the algorithms for Catfish (Human) w.r.t. number of edges.}
\label{fig:catFish3}
\end{figure}

\begin{figure}[H]
\begin{subfigure}[t]{0.45\textwidth}
    \includegraphics[trim=5 7 10 7, clip,scale=.25]{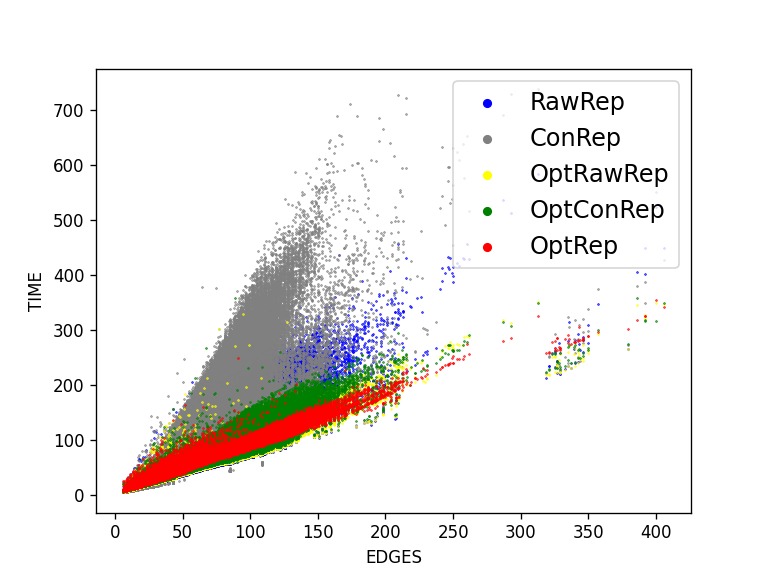}
    \caption{Time ($ms$)}
    \label{fig:timeVnode-salmon}
\end{subfigure}
\hfill
\begin{subfigure}[t]{0.45\textwidth}
    \includegraphics[trim=5 7 10 7, clip,scale=.25]{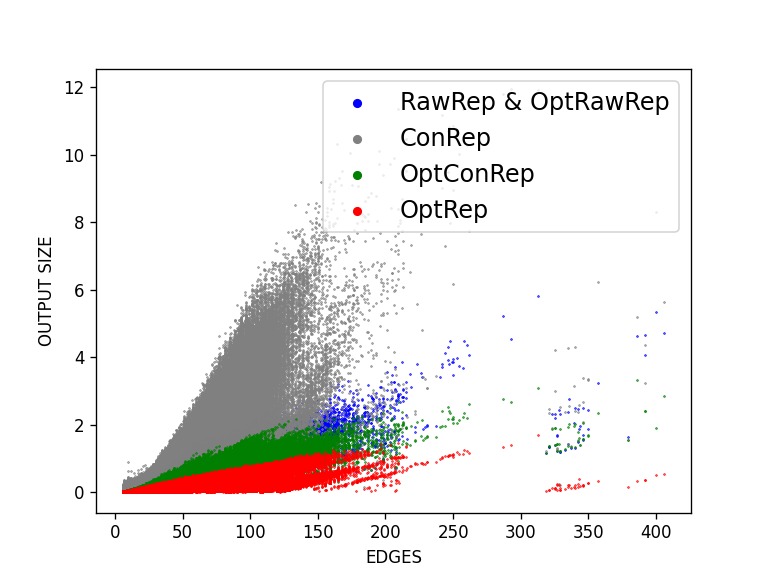}
    \caption{Output Size ($MB$)}
    \label{fig:sizeVnode-salmon}
\end{subfigure}
\caption{Performance measures of the algorithms for Catfish (Salmon) w.r.t. edges.}
\label{fig:catFish4}
\end{figure}

In the \Cref{fig:plotVnodes}, all algorithms follow an upward trend around the 1000 mark, which is surprising and not present in the plots versus the number of edges. Figure~\ref{fig:edge-vs-nodes} is a scatter plot for the edges versus nodes in the dataset. Clearly, the density of graphs is very high around the 1000 nodes mark, which explains the upward trend followed by the algorithms. The plots demonstrating performance measures with respect to nodes of the heuristic algorithms are presented \Cref{fig:allVnodesNew}.
The plots for other datasets do not show any new inferences as shown for RefSim (\Cref{fig:RefSim}) and Catfish (Overall \Cref{fig:catFish} and species \Cref{fig:catFish1,fig:catFish2,fig:catFish3,fig:catFish4}). Memory taken by individual graphs in Catfish is not shown as graphs are very small for any meaningful evaluation of memory requirements.

\end{document}